\documentclass[onecolumn]{aastex62}
\usepackage{amsmath}
\usepackage{calrsfs}
\usepackage{ulem}
\received{January 1, 2018}
\revised{January 7, 2018}
\accepted{\today}
 
\submitjournal{ApJ}

\shorttitle{Three-phase gas grain chemistry in disks} 
\shortauthors{Ruaud \& Gorti}

\begin{document}

\title{A three-phase approach to grain surface chemistry in protoplanetary disks: Gas, ice surfaces and ice mantles of dust grains}

\correspondingauthor{Maxime Ruaud}
\email{maxime.ruaud@nasa.gov}

\author[0000-0003-0522-5789]{Maxime Ruaud} 
\affiliation{NASA Ames Research Center, Moffett Field, CA, USA}

\author{Uma Gorti}
\affiliation{NASA Ames Research Center, Moffett Field, CA, USA}
\affiliation{Carl Sagan Center, SETI Institute, Mountain View, CA, USA} 

\begin{abstract}
We study the effects of grain surface reactions on the chemistry of protoplanetary disks where gas, ice surface layers and icy mantles of dust grains are considered as three distinct phases. Gas phase and grain surface chemistry is found to be mainly driven by photo-reactions and dust temperature gradients. The icy disk interior has three distinct chemical regions: (i) the inner midplane with low FUV fluxes and warm dust ($\gtrsim 15$K) that lead to the formation of complex organic molecules, (ii) the outer midplane with higher FUV from the ISM and cold dust where hydrogenation reactions dominate and, (iii) a molecular layer above the midplane but below the water condensation front where photodissociation of ices affects gas phase compositions. Some common radicals, e.g., CN and C$_2$H, exhibit a two-layered vertical structure and are abundant near the CO photodissociation front and near the water condensation front. The 3-phase approximation in general leads to lower vertical column densities than 2-phase models for many gas-phase molecules due to reduced desorption, e.g., H$_2$O, CO$_2$, HCN and HCOOH decrease by $\sim$ two orders of magnitude. Finally, we find that many observed gas phase species originate near the water condensation front;  photo-processes determine their column densities which do not vary significantly with key disk properties such as mass and dust/gas ratio.
\end{abstract}

\section{Introduction}

Protoplanetary disks consist of gas, predominantly molecular hydrogen and helium with trace amounts of CO and other molecules, and dust particles \citep[e.g.][]{Henning13,Dutrey14}. Ring-like structures observed in the radial distribution of molecules, such as $\rm N_2H^+$ \citep[e.g.][]{Qi13}, CN \citep[e.g.][]{Cazzoletti18}, $\rm C_2H$ \citep[e.g.][]{Bergin16,Kastner18}, and $\rm H_2CO$ \citep[e.g.][]{Qi13a} suggest an active coupling between gas and dust for at least some of these molecules. For example, $\rm N_2H^+$ is expected to be present where CO freezes-out at the surface of the grains, and therefore exhibit a ring-like structure; \citet{Qi13} thus indirectly derived the location of the CO snowline in the disk around TW Hya. ALMA has also revealed the presence of two complex organic molecules $\rm CH_3CN$ and $\rm CH_3OH$ ($\rm CH_3CN$ in several sources, \citet{Bergner18}, and $\rm CH_3OH$ only in one disk around TW Hya, \citet{Walsh16}). Methanol, in particular, is indicative of grain surface chemistry because it has no known efficient gas-phase formation route \citep{Geppert06,Garrod06}.
An active gas-grain chemistry has also been invoked to explain the observation of $\rm CH_3CN$ for which gas-phase chemistry alone fails in reproducing observed abundances \citep{Loomis18}. Understanding the chemistry at the surface of these grains is critical, not only to better interpret molecular observations and infer protoplanetary disk conditions but also to understand the chemical composition of the solid material from which planetary objects form \citep{Marboeuf14,Mordasini16}.

Over the years, many sophisticated models have been built to study the physics and chemistry of protoplanetary disks. However, most of these models treat the chemistry at the surface of grains in a simple manner. In many cases, the chemistry at the surface only accounts for adsorption and desorption of molecules from the surface \citep[e.g.][]{Woitke09,Agundez18}, although some include a reduced set of reactions for the formation of species such as $\rm H_2O$, $\rm CH_4$ and $\rm NH_3$ \citep[e.g.][]{Fogel11,Bruderer12,Kamp13,Cleeves14}. Only a few studies have addressed the chemistry occurring at the surface in detail by using large surface reaction networks \citep[e.g.][]{Semenov11,Walsh14}. 

Most of the models however rely on simple approximations to treat grain chemistry; often, no distinction is made between the outermost and inner monolayers of the ice; the surface and the mantle are viewed as a single phase with the same properties for diffusion, reaction and desorption (i.e.\   often referred as the 2-phase approximation and originally developed by \citet{Hasegawa92}, with one phase being the gas and the other being the ice).
One big issue with the 2-phase approach is that the desorption of mantle species, which should be remain trapped, can be considerably over-estimated.
In environments like disks, which are characterized by strong radial and vertical temperature gradients, the 2-phase approximation can strongly impact inferred disk chemical composition. 

Improvements to the 2-phase approximation have been considered, for e.g., \citet{Woitke09}, \citet{Kamp13} and \citet{Agundez18} restrict active chemistry to the top few monolayers of the ice and assume an inert mantle. However, grain chemistry in these models often only accounts for a limited number of processes (e.g.\ adsorption and desorption of molecules from the gas) and reactions \citep[e.g. formation of water by hydrogenation reactions in the chemically active surface layers;][]{Kamp13}.
The combined efforts of theoretical chemistry and experiments on interstellar dust grains have shown that some sub-surface processes (such as photodissociation, diffusion and recombination) should be at work on interstellar grain mantles \citep[e.g.][]{Andersson08,Oberg09c}. Based on these studies, chemical models used to study gas-grain chemistry in the interstellar medium have been refined to greater complexity, making a clear distinction between the chemistry occurring at the surface and in the mantle \citep[e.g.][]{Garrod13,Ruaud16,Vasyunin17}. With the exception of a few studies \citep[see][]{Drozdovskaya16,Taquet16,Yoneda16,Furuya17, Wakelam19}, the incorporation of these sophisticated models, known as 3-phase models (for the chemistry occurring in the gas phase, at the ice surface of dust grains and in the mantle) is however still largely missing in protoplanetary disk models.

In this work, we study the effects of gas grain chemistry in protoplanetary disks using a 3-phase approximation. We build on previous efforts where we developed general models for the thermochemical structure of disks \citep{Gorti08, Hollenbach09b, Gorti11}. This earlier work focused on an accurate computation of the disk thermal and vertical density structure in order to interpret surface line emission from gas species and calculate rates of mass loss from photoevaporative flows. While some processes such as high energy irradiation of the gas, dust evolution and accompanying changes in attenuation were studied in considerable detail, chemistry in the optically thick dusty midplane regions, where gas and dust temperatures are well-coupled, was treated very simplistically.  Important processes like condensation of molecules were in fact ignored. Here, we combine the thermochemical models with an updated, state-of-the-art 3-phase gas grain chemical code \citep{Ruaud16}. The gas grain chemical code was originally developed and used to study cloud cores in the interstellar medium, and here has been updated to a 2D formalism applicable to disks. The treatment of photo-reactions, in particular, has been substantially revised and UV radiative transfer (in discrete energy bins) has been added to comply with the thermochemical model. As we show in this paper, photo-processes dominate the chemistry of observable species in the disk, both in the gas phase and  on grain surfaces. 

The basic modeling framework is presented in this paper, and it will be used to model individual disks in the future to infer disk chemical and physical conditions. We find that the chemistry is tightly coupled to the underlying disk structure and moreover surprisingly robust, and consider a few disk models to demonstrate this result. This paper is organized as follows. In Section \ref{sec:model_description} we present the general framework of the model. In Section \ref{sec:model_results}, we present our modeling results and discuss the main differences found with the more widely used 2-phase approximation. In Section \ref{sec:discussion} compare these results with other models and discuss existing disk observations. Section \ref{sec:conclusions} summarizes this work.

\section{Model description and overview}
\label{sec:model_description}

We first compute the physical structure of the disk using a thermochemical model (described in detail in appendix \ref{sec:thermochem}) and then use the physical structure to determine disk gas-grain chemistry (in an approach similar to \citet{Walsh12}).

The main inputs to the thermochemical model are the gas and dust surface density distributions, and stellar parameters such as mass and optical/UV/X-ray spectrum. We solve for vertical hydrostatic pressure equilibrium, coupled iteratively with gas heating/cooling and chemistry. Dust grains follow a power law size distribution ($n_d(a) \propto a^{-3.5}, 0.005 \mu\mathrm{m} < a < 1 \mathrm{cm})$ and are composed of silicates and carbon grains. Dust grains settle vertically with a scale height set by gas turbulence, and we adopt a value for the viscous parameter, $\alpha=0.005$. Many iterations are needed to solve for a self-consistent disk physical structure (density and temperature as functions of $(r,z)$), and in order to keep the problem computationally tractable we solve for steady state chemistry while computing the gas line cooling (nLTE using escape probabilities). We also do not integrate the gas-grain chemical network directly into the thermo-chemical model for the same reason. In the present framework, the equilibrium gas physical structure (i.e.\  gas density and temperature) and the dust physical structure (i.e.\  moments of the grain size distribution, dust temperature and the $A_V$) are used as inputs of the time dependent gas-grain chemical model. We post-check after the chemical simulations to ensure that the gas phase abundances of the main cooling species are consistent between both models. 

The gas-grain chemical model is based on the time-dependent 3-phase gas-grain chemical model presented in \citet{Ruaud16}. This model uses the rate equation approach to solve the gas-phase and the grain surface chemistries and obtain the fractional abundance of each chemical species at each location of the disk as a function of time. The gas-phase and the grain surface chemistry are fully coupled by accretion and desorption processes. Grain surface chemistry includes diffusion of the chemical species on the surface, 2-body reactions, photo-reactions, as well as thermal and non-thermal desorption mechanisms (see \S \ref{sec:grain_chem} and Appendix \ref{sec:grain_chemistry} for a detailed description). 
In this 3-phase approach, we treat the chemistry occurring at the surface and in the bulk of the ice as two distinct phases in interaction. As in \citet{Ruaud16}, we account for surface roughness by assuming that the surface consists of two outermost layers, rather than just the top layer. Once sufficient material has accreted onto the surface (i.e.\  all the surface adsorption sites are saturated), the mantle layers begin to be populated by incorporating material from the surface. Like the surface, the mantle material is considered to be chemically active and subject to photodissociation, diffusion and reaction. Desorption is restricted to the surface layers. In the mantle, where water ice is the main constituent of the ice, we assume that diffusion is the result of a collective molecular motion rather than an individual process and thus assume that molecular diffusion is driven by self diffusion of water molecules \footnote{Note that this treatment allows the model to reproduce observed desorption kinetics by TPD (Temperature Programmed Desorption) experiments, for which first- and zero-order desorption kinetics are at work, relatively well. See Appendix A in \citet{Ruaud16}.} (see Appendix \ref{sec:grain_chemistry}). It is important to note that in the 3-phase approximation, the net rate of change of the total surface and mantle material is not due to a true physical motion of material from the surface to the mantle (and vice versa in case of net desorption) but due to the continuous renewal of the grain surface material. True physical motion is taken into account by a swapping process in which molecules from the surface can swap their position with molecules from the bulk and {\it vice versa}. 

\subsection{General equations}
\label{sec:general_eq}
In the following, gaseous, surface and mantle concentrations of a species $i$ are denoted as $n(i)$, $n_s(i)$ and $n_m(i)$ and have unit of cm$^{-3}$. In some places, we use the notation $n_\text{ice}(i)$ which stands for both; surface species and mantle species. The same notation is used for surface and mantle processes.
The differential equations governing $n(i)$, $n_s(i)$ and $n_m(i)$ are given by \citep{Ruaud16}: 

\begin{eqnarray}
\frac{dn(i)}{dt}\biggr\rvert _\text{tot} &=&\sum_l \sum_j k_{lj}n(l)n(j) + k_\text{ph}(j)n(j) \nonumber \\
& &+ k_\text{des}(i)n_s(i) - k_\text{acc}(i)n(i) \nonumber \\
& & - k_\text{ph}(i)n(i) - n(i)\sum_l k_{ij} n(j)
\end{eqnarray}
\begin{eqnarray}
\frac{dn_s(i)}{dt}\biggr\rvert _\text{tot}&=&\sum_l \sum_j k_{lj}^sn_s(l)n_s(j) + k_\text{ph}^s(j)n_s(j) \nonumber \\
& &+ k_\text{acc}(i)n(i) + k_\text{swap}^{m}(i)n_m(i) \nonumber \\
& & + \frac{dn_m(i)}{dt} \biggr\rvert _{m\rightarrow s}  - n_s(i)\sum_j k_{ij}^s n_s(j) \nonumber \\
& & -  k_\text{des}(i)n_s(i) - k_\text{ph}^s(i)n_s(i)\nonumber \\
& & - k_\text{swap}^{s}(i) n_s(i) - \frac{dn_s(i)}{dt} \biggr\rvert _{s\rightarrow m}
\end{eqnarray}
\begin{eqnarray}
\frac{dn_m(i)}{dt}\biggr\rvert _\text{tot}&= &\sum_l \sum_j k_{lj}^m n_m(l)n_m(j) + k_\text{ph}^m(j)n_m(j) \nonumber \\ 
& &+ k_\text{swap}^{m}(i) n_s(i)  + \frac{dn_s(i)}{dt} \biggr\rvert _{s\rightarrow m} \nonumber \\
& &- n_m(i)\sum_j k_{ij}^m n_m(j) - k_\text{ph}^m(i)n_m(i) \nonumber \\
& & - k_\text{swap}^{m}(i) n_m(i) - \frac{dn_m(i)}{dt} \biggr\rvert _{m\rightarrow s}
\end{eqnarray}
$k_{ij}$, $k_{ij}^s$ and $k_{ij}^m$ [cm$^3$s$^{-1}$] represent the rate coefficients for reactions between species $i$ and $j$ in the gas-phase, on the grain surface and in the mantle respectively. $k_\text{acc}$ and $k_\text{des}$ [s$^{-1}$] are respectively the accretion/evaporation rate coefficient of individual species onto/from the grain surface. In our model $k_\text{des}$ can be thermal or non-thermal. $k_\text{ph}$ [s$^{-1}$] collectively denotes dissociation and ionization rate coefficients by (1) FUV and X-ray photons, (2) cosmic-rays and (3) secondary UV photons induced by cosmic-rays and X-rays. $k^s_\text{swap}$ and $k^m_\text{swap}$ [s$^{-1}$] are the surface-mantle and mantle-surface swapping rate coefficients, respectively. $dn_s(i)/dt\rvert_{s\rightarrow m}$ and $dn_m(i)/dt\rvert_{m\rightarrow s}$ are related to the individual transfer of the species $i$ from the surface to the mantle and \textit{vice versa}. 
The net rate of change at the surface depends on adsorption, desorption and also on chemical reactions \citep[see][]{Ruaud16}. We note that our approach slightly differs from the one originally proposed by \citet{Hasegawa93}, in that in our formulation, a molecule lost at the surface is immediately replaced by transfer of a molecule from the mantle to the surface \citep[see also][]{Ruaud16}.

\subsection{Gas-phase chemistry}
We have made some modifications to the gas phase network of \citet{Ruaud16} which was based on the {\it kida.uva.2014} network \citep{Wakelam15}. Several processes such as X-ray photoreactions, reactions with PAHs as well as reactions with excited H$_2$ (denoted as H$_2^*$ in the following) have been added to this network (see Appendix \ref{sec:gas_chemistry}). For  ionisation by cosmic rays we follow \citet{Padovani18} who provide a simple analytical expression as a function of the gas column density (see Appendix \ref{sec:cosmicray}). 

In protoplanetary disks, dust evolution (settling, coagulation) results in a reduction of FUV photon attenuation \citep[e.g.\ ][]{Gorti15}. Hence the opacity due to different atoms and molecules, from both the gas and the ice, may become significant for FUV shielding. Therefore, the treatment of the photoprocesses has been modified to comply with the thermochemical models and uses a 1+1D approximation which includes a computation of the FUV (and optical) photon absorption by the atoms and molecules (by gas and ice). We follow the method described in \citet{Gorti04} and split the UV-visible region into nine energy bins from 0.74 to 13.6 eV (H$^-$, CH and CH$^+$ are species that could potentially contribute to the total molecular opacity at visible-NUV energy bands ($0.74-6$ eV)). The nine intervals were chosen to correspond to dominant gas photoabsorption thresholds and include 0.74-2.6 eV, 2.6-3.5 eV, 3.5-4.3 eV, 4.3-5.12 eV, 5.12-7.5 eV, 7.5-10.0 eV, 10.0-10.36 eV, 10.36-11.26 eV and finally 11.26-13.6 eV. When data are available, {gas-phase} photoabsorption, photodissociation and photoionisation cross sections are integrated over each energy bin to get the weighted cross sections for each species in each bin (see Appendix \ref{sec:gas_chemistry}). 
We choose to not make a distinction between the FUV opacities of gas and condensed phases. Although absorption cross sections have been measured for a few ice species \citep[e.g.][]{Mason06,CruzDiaz14a,CruzDiaz14},  these measurements cover a limited range of the FUV spectrum with no data at higher ($\sim 10.36$ eV) energies. Where available, data indicate that the weighted cross-sections in our FUV bins are not significantly different for gas and ice; e.g, the integrated cross-sections  in the 7.5-10.0 eV range for H$_2$O and CO$_2$ are almost identical for the gas and condensed phases. The cross-section of CO$_2$ in the Ly-$\alpha$ energy bin (10-10.36 eV) is one notable exception, where ice measurements are a factor of $\sim 15$ greater than gas-phase CO$_2$. We note that there is considerable uncertainty in the indirectly derived stellar Ly-$\alpha$ flux (e.g., France et al. 2014), and that other parameters, such as the temperarure of the ice, can also impact the absorption cross-sections by a factor of few \citep[e.g. see][]{Mason06}. We therefore use the same opacities for both gas and condensed phases for all species and hereafter refer to these collectively as molecular opacity.

The network used by the thermo-chemical model \citep{Gorti11} differs from the {\it kida.uva.2014} network, but to maintain consistency between the thermal structure obtained and the gas-grain chemical modeling we check that the results obtained by the two models show good agreement for most of the abundant species in the disk. We especially ensure that the main coolant abundances agree, this includes most of the metallic atoms and ions as well as molecules like H$_2$, CH, CO, OH, H$_2$O and HCO$^+$. These species dominate gas cooling mostly in the surface layers of disks where photo-processes are important, and the same treatment of photo-rates has been used for both models. Consequently, although our model setup has been devised to allow iterations between the two models if necessary, we find that the model abundances agree to $\lesssim 5\%$ in relevant regions even without any iterations. The exceptions are CN, HCN and C$_2$H, which are not included in the reduced network but may contribute to cooling as they have been observed in emission from protoplanetary disks. We compute the expected line cooling from the abundances obtained from the full chemical network and find that although these lines can be bright, their contribution to total gas cooling is at most $\sim 10$\%. Since the cooling function depends only weakly on temperature in relevant regions, the change in gas temperature is expected to be even smaller.  We therefore neglect the cooling from these species in this paper. 

\subsection{Grain surface chemistry}
\label{sec:grain_chem}

Grain surface chemistry includes accretion of atoms and molecules at the surface of the grains, diffusion, reaction, photoprocesses as well as thermal and non-thermal desorption mechanisms, i.e\ photodesorption and chemical desorption (see Appendix \ref{sec:grain_chemistry}). In the following we use a constant diffusion-to-binding energy ratio of 0.4 for surface species and 0.8 for the mantle and  a contant photodesorption yield of $10^{-4}$ molecules per photon (see Appendix \ref{sec:grain_chemistry} for more information).

We consider a distribution of sizes for the grains and solve for the grain surface chemistry by using size-integrated quantities at each $(r,z)$ location of the disk.
We compute the effective ice thickness for a size distribution and accordingly consider growth of grains by condensation of material from the gas. 
To calculate the ice thickness we follow an approach similar to \citet{Guillet07}. The total mass density of the ice averaged over the whole mantle is given by
\begin{equation} \label{eq:guillet}
    \int_a \frac{4}{3}\pi \rho \Big( (a+\Delta a)^3 - a^3\Big) dn_d(a) = \rho_\text{ice}
\end{equation}
where $\rho$ is the average material density of the ice material and taken to be 1 g cm$^{-3}$, $a$ the grain radius, $\Delta a$ ice thickness and $\rho_\text{ice} = \sum n_\text{ice}(i)m(i)$ the total mass density of the ice summed over all the species present in the ice. 
From the $k$-th order moments of the grain size distribution
\begin{equation}
\label{eq:moments}
    \langle a^k n_d \rangle = \int_a a^k dn_d(a)
\end{equation}
the ice thickness $\Delta a$ is the unique, real solution of the cubic equation
\begin{equation}
    \langle n_d \rangle \Delta a^3 + 3 \langle an_d \rangle\Delta a^2 + 3 \langle a^2 n_d \rangle \Delta a = \frac{3}{4 \pi} \frac{\rho_\text{ice}}{\rho}
\end{equation}
We solve for $\Delta a$, and then find the total number of monolayers at the surface of the grains, $N_\text{lay} = \Delta a / d_\text{sites}$, where $d_\text{sites}$ is the mean distance between two adsorption sites and assumed to be equal to 2.6\AA\  \citep{Hasegawa92}.  The number density of adsorption sites per monolayer, $n_\text{sites}$, is then computed by
\begin{equation}
\label{eq:nsites}
    n_\text{sites} = \frac{\sum{n_\text{ice}(i)}}{N_\text{lay}}
\end{equation}
where the summation is made over all the species present in the ice. By doing so, we ensure that the number of monolayers present at the surface satisfy the mass conservation equation given in Eq. \ref{eq:guillet}. 
The product of the dust cross sectional area and the dust density averaged over the grain size distribution is then computed by
\begin{equation}
    \langle \sigma_d n_d \rangle = \int_a \pi (a+\Delta a)^2 dn_d(a)
\end{equation}
which reduces to (by using Eq. \ref{eq:moments})
\begin{equation}
\label{eq:geo_cross_sec2}
    \langle \sigma_d n_d \rangle = \pi\langle a^2 n_d \rangle + 2\pi\Delta a \langle a n_d \rangle + \pi\Delta a^2 \langle n_d \rangle
\end{equation}
This expression takes into account the impact of the growth of the grain by condensation of molecules from the gas on the dust cross sectional area.

\section{Model results}
\label{sec:model_results}

In the following we consider a series of different disk models where we vary the disk physical parameters. We assume $M_*=1M_\odot$, $R_* = 2.45 R_\odot$ and an X-ray luminosity $L_X = 10^{30}$ erg/s. The stellar spectrum corresponds to that of TW Hya  \citep{Heays17} as available in the Leiden database (LAMDA)\footnote{\url{http://home.strw.leidenuniv.nl/~ewine/photo/}}. The X-ray spectrum is a 2-temperature plasma with components at 5 and 10MK. The FUV component of the interstellar radiation field is approximated by the expression given in \citet{Sternberg95}, which provides a fit to the radiation field of \citet{Draine78}. We hereafter define $G_0$ as the total integrated photon flux (over the $6-13.6$ eV range) at a given location, normalized to the Habing field \citep{Habing68}, and use this as a measure of the total local FUV field. The density and temperature structure of the disk is first calculated by the thermo-chemical model and assumed to be static while solving for the time-dependent gas-grain chemistry. As in the thermo-chemical code, the photorates are self consistently recalculated by the gas-grain chemical model at each time step as discussed in Section \ref{sec:photoprocesses}. This includes the computation of the column densities $N_i(r,z)$ of each absorber species $i$ and the resulting attenuation factors $\tau_n(r,z)$ in each energy bin.

The initial chemical conditions in the disk are determined by evolving a dark cloud model \citep[see][]{Ruaud16} for 1Myr with constant physical parameters ($n_\text{H}=2\times10^4$ cm$^{-3}$, $T_\text{g}=T_\text{d}=10$ K, $A_V=10$ mag) and initial elemental abundances as given in Table \ref{tab:init_ab}. The fractional abundances thereby obtained
(also given in Table \ref{tab:init_ab}) are used as inputs for the time-dependent chemical evolution of the disk model. Except where explicitly mentioned, all the results are shown after 1 Myr of disk evolution. 
In the following, molecules denoted as $sX$ refer to molecules present in the ice (total fractional abundance includes surface and mantle components). We note that for most of our models, chemistry is not sensitive to the initial chemical composition and that a chemical reset takes place in most of the disk due to photoprocesses. One exception is the inner disk midplane, where the ionisation is dominated by cosmic rays (slow chemical processes). This will be further discussed in Section \ref{sec:general_results_3}.

\begin{table}
\centering
\caption{\label{tab:init_ab} Adopted elemental abundances and initial disk fractional composition. The initial disk fractional composition was obtained from the results of a time-dependent dark cloud model with contant physical parameters, i.e.\ : $n_\text{H}=2\times10^4$cm$^{-3}$, $T_g=T_d=10$K, $A_V=10$ mag, and after 1 Myr evolution. Molecules denoted as s$X$ refer to molecules present at the surface of dust grains.}
\begin{tabular}{lclc}
\hline
\hline
Element 	& Abundance & Element & Abundance \\
\hline
H		& 1.0 	      & He 	  & 0.1 \\
O		& 3.2(-4)	& C & 1.4(-4) \\
N		& 6.2(-5)	& Si & 1.7(-6) \\
Mg 		& 1.1(-6)	& S & 1.0(-6) \\
Fe		& 1.7(-7)	& PAH & 1.0(-9) \\
\hline
Species & Abundance & Species & Abundance \\
\hline
$\rm H_2 $          &   0.5 & $\rm He $           &   0.1 \\
$\rm CO $          &   9.4(-6) &$\rm H_2O $          &   4.0(-8) \\
$\rm N_2 $          &   1.0(-6) & $\rm S $          &  1.0(-7) \\
$\rm Mg^+ $       &   8.0(-8) & $\rm Mg $       &   7.0(-8)\\
$\rm Si $       &   5.0(-8) & $\rm SiO$       &   5.0(-8)\\
$\rm Fe^+ $       &   5.0(-8) & $\rm Fe$       &   2.0(-8)\\
$\rm sH_2O$         &   1.8(-4) & $\rm sCO $          &   5.8(-5)\\
$\rm sCO_2 $        &   1.8(-5) & $\rm sN_2 $         &   1.4(-5) \\
$\rm sCH_3NH_2 $    &   1.2(-5) & $\rm sHCN $         &   9.6(-6) \\
$\rm sCH_4 $        &   6.7(-6)  & $\rm sHCOOH $       &   6.2(-6) \\
$\rm sCH_3OH $      &   4.6(-6) & $\rm sNH_3 $        &   4.1(-6) \\
$\rm sNO $          &   2.3(-6) & $\rm sHNO $         &   2.0(-6) \\
$\rm sCH_3OCH_3 $   &   1.8(-6) & $\rm sC_2H_6 $      &   1.6(-6) \\
$\rm sH_2CO $      &   1.2(-6) & $\rm sSiH4 $   &   1.2(-6) \\
$\rm sMgH_2 $   &   9.5(-7) & $\rm sCH_3CCH $   &   8.7(-7) \\
$\rm sSiO $   &   4.0(-7) & $\rm sHS $          &   2.1(-7)\\
$\rm sH_2S $        &   2.0(-7) &   $\rm sSO $          &   2.0(-7) \\
$\rm sH_2CS $       &   1.2(-7) &   $\rm sOCS $          &   1.0(-7) \\
$\rm sFeH $       &   1.0(-7) & $\rm sSO_2 $       &   1.8(-8)\\
\hline
\end{tabular}
\tablecomments{$a(b)$ stands for $a\times10^b$}
\end{table}

\subsection{General results}
We first discuss the results of the fiducial disk model ($M_\text{disk}=0.005M_\odot$, $\sim$ mean disk mass from the Taurus  disk survey \citep{Andrews05}). The disk extends from 1 to 200 au and the gas and dust surface density distributions follow a $1/r$ power law with radius. The dust-to-gas mass ratio is set to  $\Sigma_\text{d}/\Sigma_\text{g}=0.01$ and is constant with radius, but varies with height due to dust settling. Dust grains have a range of size distributions (32 bins) and settle vertically with a scale height set by the local gas density, temperature and turbulence. In the following, we discuss the chemical structure of the disk and focus on molecules that are either the most abundant or have been observed in disks, and discuss their associated gas-phase and grain surface chemistry. We first provide an overview of the physical and chemical structure (\S \ref{sec:general_results_1}) and then describe the chemistry in the molecular layer (\S \ref{sec:general_results_2}), and in the disk midplane (\S \ref{sec:general_results_3}) for the 3-phase model. We note that most chemical pathways for 3-phase and 2-phase models are similar and differences are mainly due to relative efficiencies of the pathways.  Readers interested only in a comparison with the 2-phase model may wish to skip here to  \S \ref{sec:2vs3phase}.
Since the focus of this work is on gas-grain chemistry which is not very important at the disk surface, we do not discuss the chemistry of the atomic and ionized layers at the surface and refer the reader to earlier work \citep{Gorti08, Gorti11} .

\subsubsection{Physical and chemical disk structure overview}
\label{sec:general_results_1}

\begin{figure*}
\centering
\includegraphics[width=18cm,trim=0 0.7cm 0 0,clip]{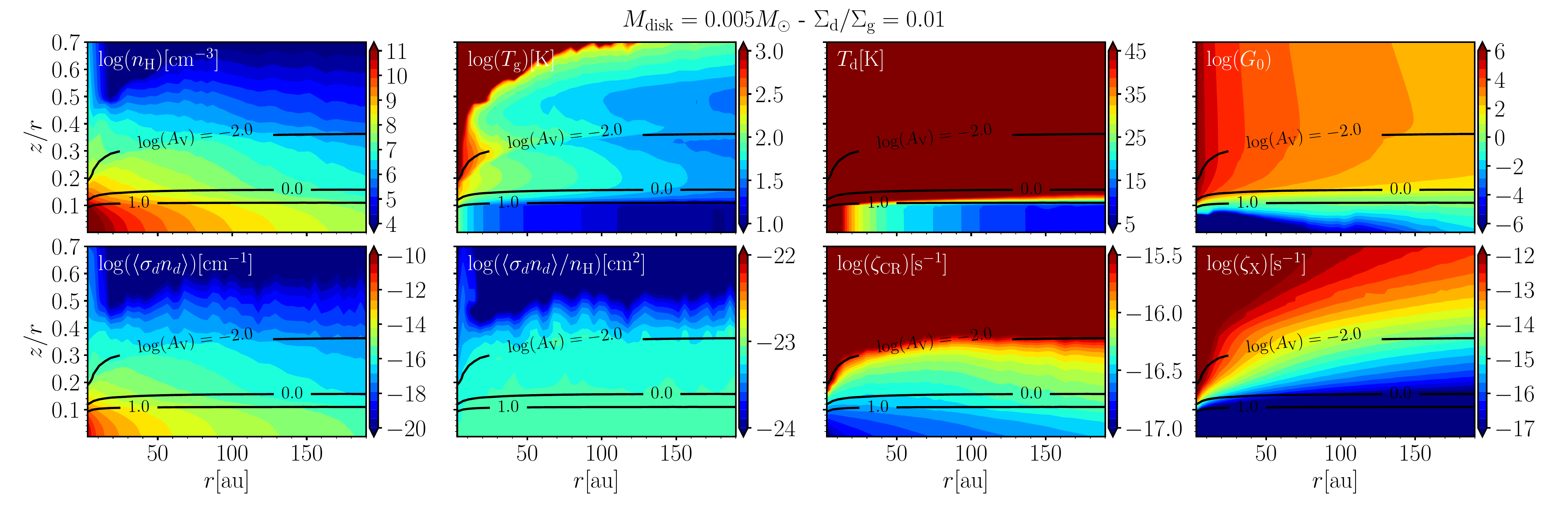}
\caption{\label{fig:default_disk_struc} Physical structure of the fiducial disk  computed with the thermo-chemical model. The disk is assumed to have a mass $M_\text{disk}=0.005M_\odot$ and the dust-to-gas mass ratio is set to $\Sigma_\text{d}/\Sigma_\text{g}=0.01$. The gas density ($n_\text{H}$), gas temperature ($T_\text{g}$), dust temperature ($T_\text{d}$),  the local stellar and interstellar FUV field ($G_0$), the product of the dust cross sectional area and the dust density averaged over the grain size distribution ($\langle \sigma_\text{d}n_\text{d} \rangle$), the cross-sectional grain area per H atoms ($\langle \sigma_\text{d}n_\text{d} \rangle / n_\text{H}$), the cosmic ray ionisation rate ($\zeta_{\text{CR}}$) and the X-ray ionisation rate ($\zeta_{\text{X}}$) are shown as a function of the disk radius $r$ and the normalized height $z/r$. Solid line contours show the visual extinction ($A_V$) to the star.}
\end{figure*}

Gas-grain chemistry depends on the key physical parameters shown in Fig.\ref{fig:default_disk_struc}: the computed gas density, gas temperature, dust temperature, mean dust cross sectional area, local (stellar and interstellar) FUV field, cosmic ray and X-ray ionisation rate. The solid line contours show the $A_V$ to the star\footnote{In what follows, we often use $A_V$ to the star (instead of $z/r$) as an indicator of disk vertical height to allow comparisons between disk models that could have diferent physical structures. Regions with a high $A_V$ from the star may therefore still be subject to high incident photon fluxes from interstellar and scattered photons.}. At $A_V\gtrsim1.0$ mag ($z/r\sim 0.15$), the FUV field begins to be attenuated and at $A_V\gtrsim3.0$ mag the FUV flux becomes low enough to allow rapid growth of ices; i.e.\ photodesorption can no longer counterbalance the formation of water ice on grain surfaces. At $r\gtrsim 100$ au, interstellar FUV photons and some scattered photons from the star (mostly Ly$\alpha$) can penetrate the midplane (Fig. \ref{fig:default_disk_struc}). Radial variations in local physical parameters  strongly influence the chemistry of both the gas and on grains. 
At $ r\gtrsim 10$ au, high densities and low temperatures in the midplane region lead to efficient condensation of most of the gas-phase molecules into thick ice mantles on grains and grain surface chemistry becomes significant. 
Closer to the star, $r\lesssim 2-10$ au, dust is too warm to form ices even in the midplane and the chemistry is dominated by gas-phase processes.

\begin{figure*}
\centering
\includegraphics[width=14cm,trim=0 0.7cm 0 0,clip]{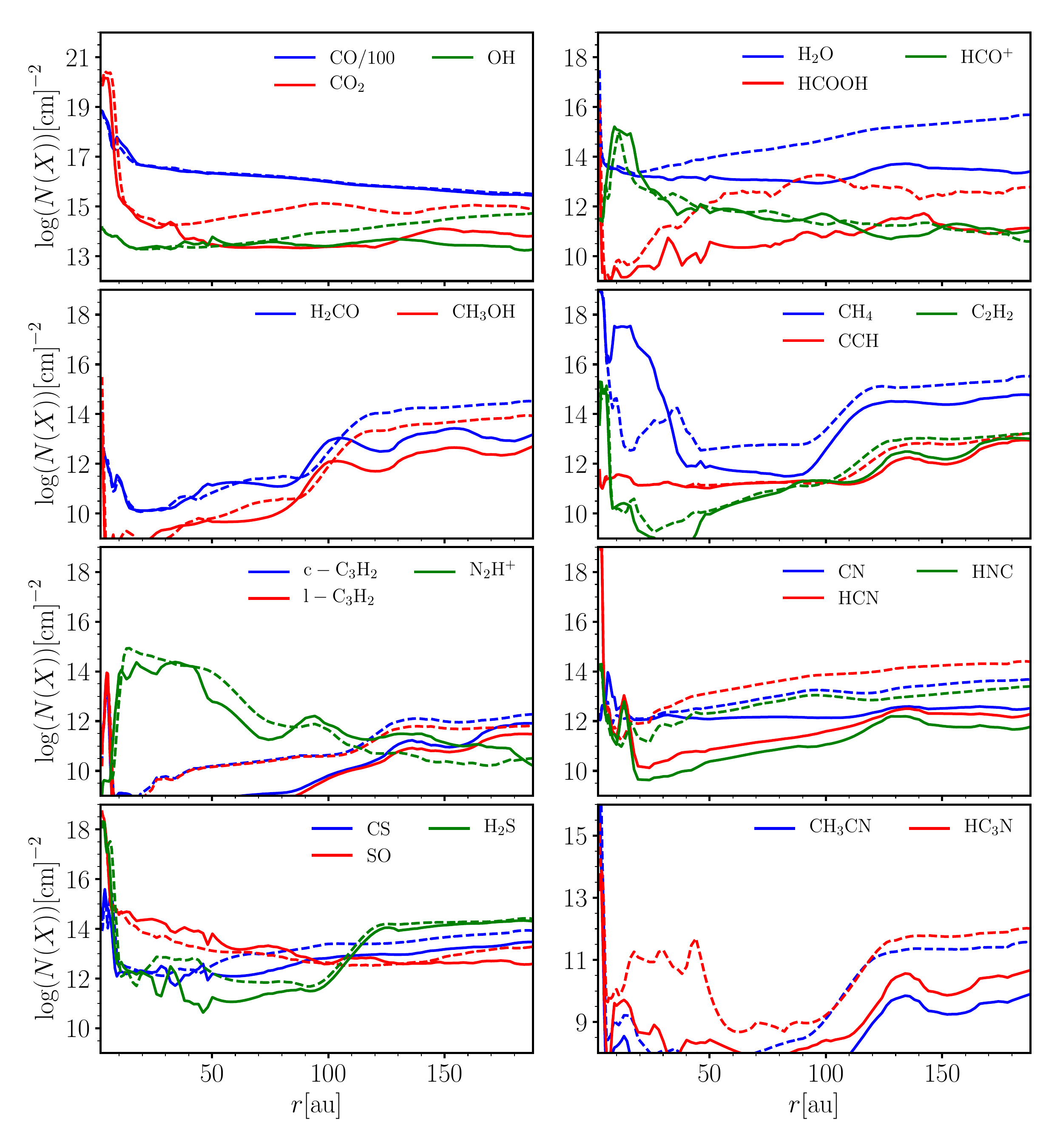}
\caption{\label{fig:col_dens_gas} Computed vertical column densities of a selection of gas-phase molecules as a function of the disk radius $r$ for the fiducial disk model (i.e.\  $M_\text{disk}=0.005M_\odot$ and $\Sigma_\text{d}/\Sigma_\text{g}=0.01$). Solid lines show the results obtained with the 3-phase model while dashed lines are those obtained from a 2-phase model (see Section \ref{sec:2vs3phase})}
\end{figure*}

\begin{figure*}
\centering
\includegraphics[width=14cm,trim=0 0.7cm 0 0,clip]{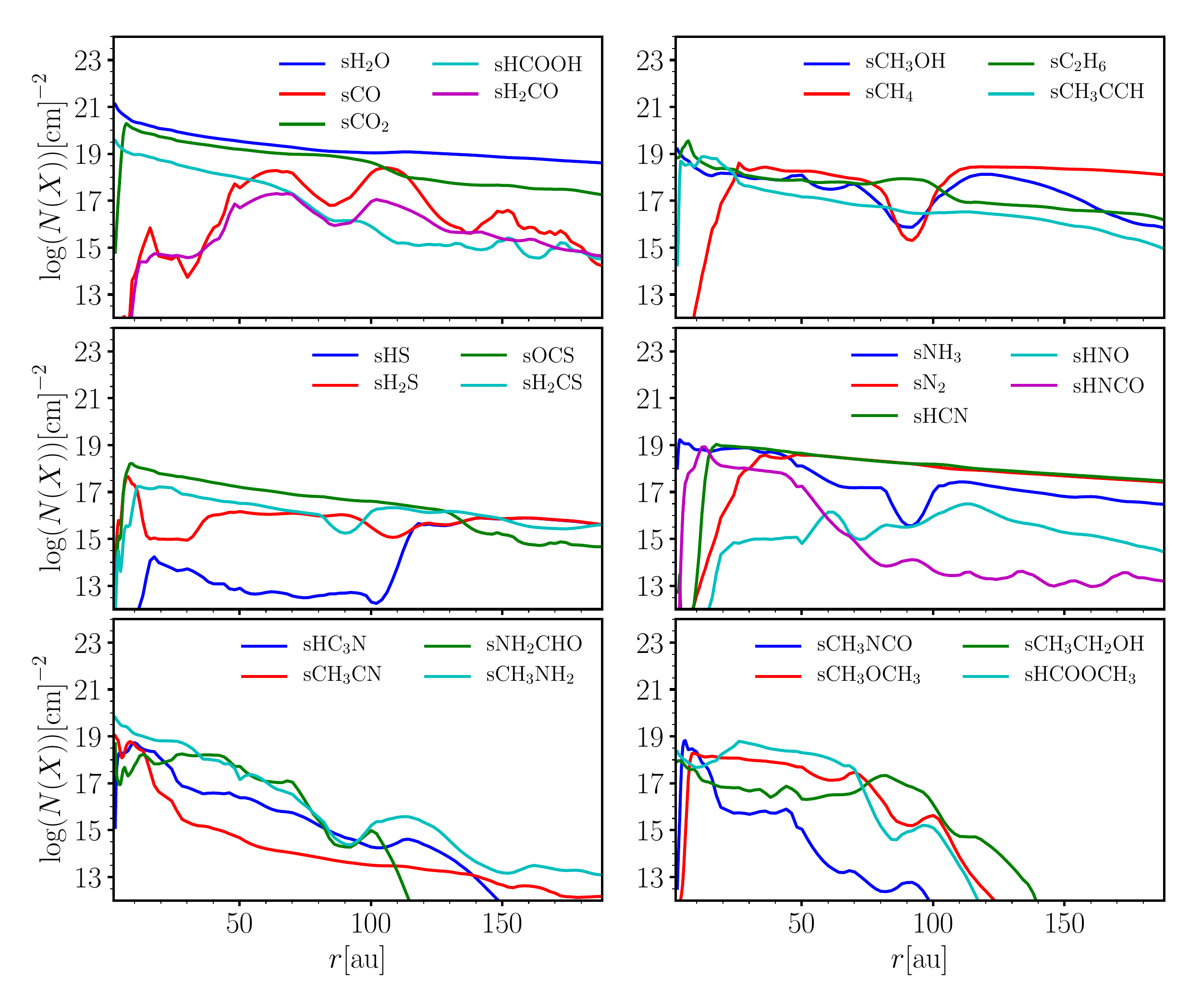}
\caption{\label{fig:col_dens_grain} Computed vertical column densities of a selection of molecular ices on grains (sum of contributions from ice surface and ice mantles) as a function of the disk radius $r$ for the fiducial disk model (i.e.\  $M_\text{disk}=0.005M_\odot$ and $\Sigma_\text{d}/\Sigma_\text{g}=0.01$).}
\end{figure*}

Figures \ref{fig:col_dens_gas} and \ref{fig:col_dens_grain} show the computed vertical column densities of a selection of gas-phase and grain surface molecules as a function of disk radius. (Note that Fig. \ref{fig:col_dens_gas} also shows 2-phase model results, but we defer comparisons to Section \ref{sec:2vs3phase}). Gas-phase molecules include all molecules observed to date in disks while grain surface molecules include the most abundant species contained in ices on grains. 
Some of the gas phase molecules show clear ring-like structures; e.g. $\rm N_2H^+$ and $\rm HCO^+$ show enhanced vertical column densities on the order of $\sim 10^{14}-10^{15}$ cm$^{-2}$ in the range $10 \lesssim  r \lesssim 50$ au and $10 \lesssim  r \lesssim 25$ au respectively (Fig. \ref{fig:col_dens_gas}). 
Clear deficits are seen at small radii for $\rm H_2CO$, $\rm CH_3OH$ and for most of the carbon chains which show enhanced vertical column densities at $r\gtrsim 100$ au. Note that when observed, such structures would appear ring-like in morphology due to the outer cut-off of $\Sigma_g$ at the outer disk edge.

Fig. \ref{fig:col_dens_grain} shows that water  dominates ice composition in most of the disk midplane followed by simple molecules such as $\rm sCO$, $\rm sCO_2$, $\rm sCH_4$, $\rm sNH_3$, sHCN, $\rm sN_2$, $\rm sCH_3OH$ and carbon chains. The inner disk ($r \lesssim 100$ au) is characterized by the presence of large quantities of complex organic molecules such as $\rm sCH_3OCH_3$, $\rm sCH_3CH_2OH$, $\rm sCH_3COCH_3$, $\rm sHCOOCH_3$, $\rm sCH_3NH_2$, $\rm sNH_2CHO$ and $\rm sCH_3NCO$, which are conspicuously absent at $r \gtrsim 100$au.

Based on the above general structure, we denote various regions of the disk as follows. The vertical zone between the H/H$_2$ transition and height $z/r$ where gas-grain chemistry dominates is called the molecular layer (see \S\ref{sec:general_results_2}, $z/r \gtrsim 0.12$, $5\times10^{-3} \lesssim A_V \lesssim 3.0$ mag). The region where $z/r\lesssim 0.12$, $A_V \gtrsim 3.0$ mag marks the inner ($r \lesssim 100$ au) and outer ($r\gtrsim100$au) disk midplanes. This distinction is motivated by a key transition in the disk chemical composition (discussed later) at $r\sim 100$au  where the dust temperature $T_d$ drops below $\sim 15$ K. Beyond $\sim$100 au, chemistry is dominated by hydrogenation reactions (i.e. reduced radical mobility and larger residence time of hydrogen at the surface), while in the inner disk midplane dust is warm enough (i.e.\  $T_d \gtrsim 15$ K ) to allow efficient diffusion of  radicals at the surface and radical reactions successfully compete with hydrogenation reactions to form large complex organic molecules on grain surfaces.

\subsubsection{Chemistry of the molecular layer}
\label{sec:general_results_2}

\begin{figure*}
\centering
\includegraphics[width=14cm,trim=0 0.7cm 0 0,clip]{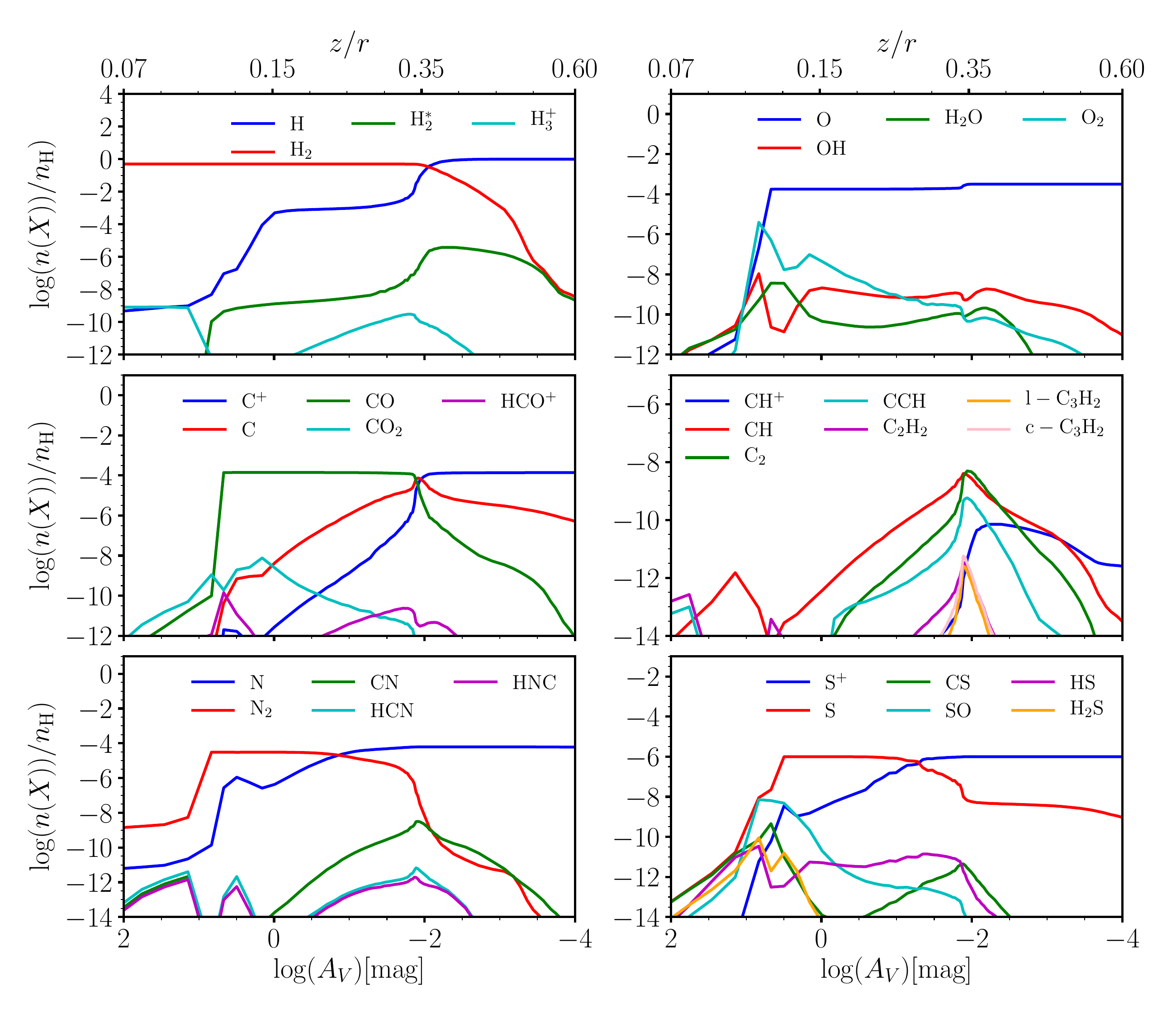}
\caption{\label{fig:cut_r100} Abundance of key gas-phase species in the molecular layer for hydrogen, oxygen, carbon, nitrogen and sulfur shown as a function of the $A_V$ from the star at $r=50$au for the fiducial disk model ($M_\text{disk}=0.005M_\odot$ and $\Sigma_\text{d}/\Sigma_\text{g}=0.01$). The drop in abundances seen at $A_V\sim10$ mag is due to the formation of ices at grain surfaces.}
\end{figure*}

\begin{figure*}
\centering
\includegraphics[width=18cm,trim=0 0.7cm 0 0,clip]{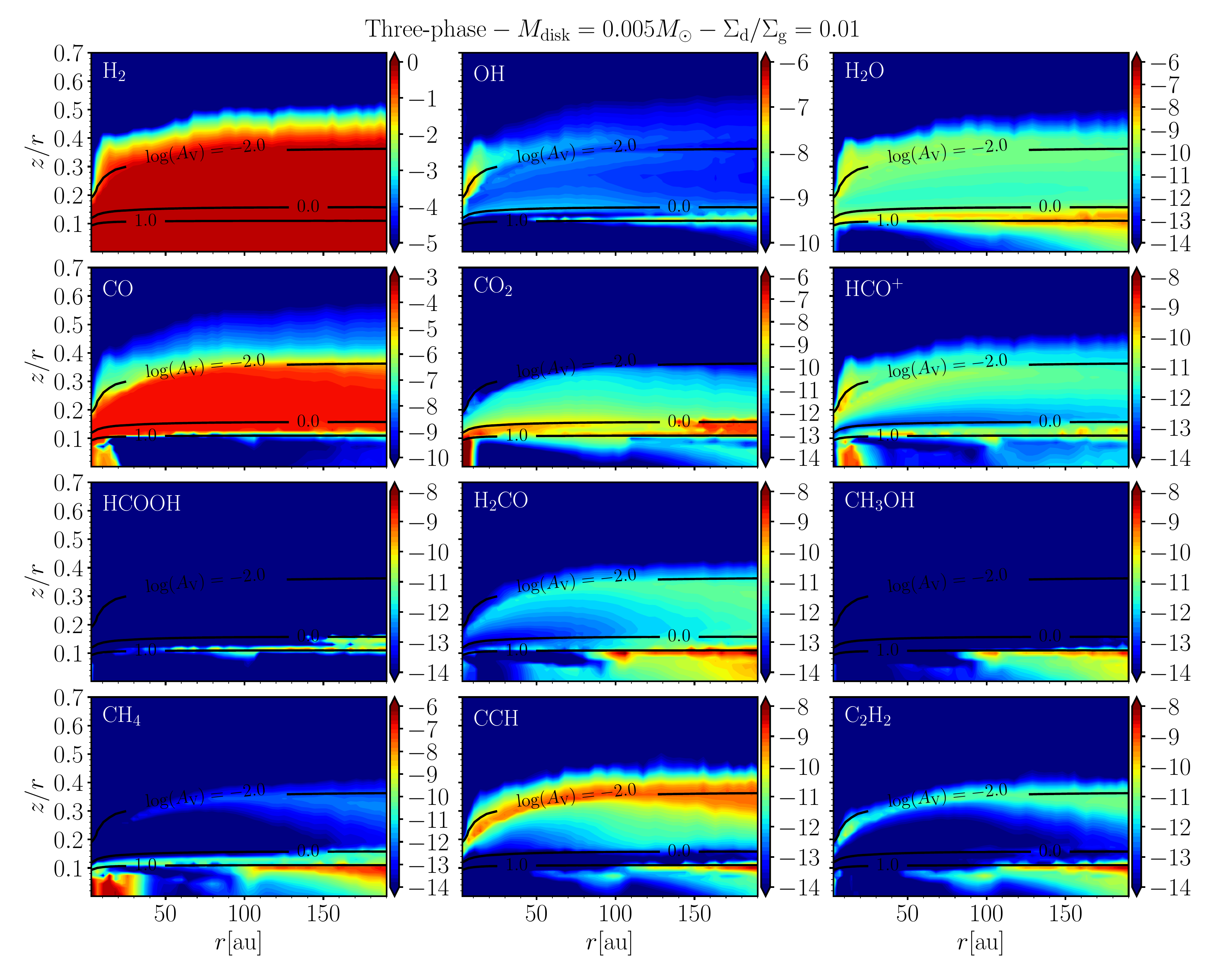}
\caption{\label{fig:ab_prof1} Computed gas-phase abundances of a selection of species as a function of the disk radius $r$, and normalized height $z/r$. Solid line contours show the $A_V$ to the star for the fiducial disk model ($M_\text{disk}=0.005M_\odot$ and $\Sigma_\text{d}/\Sigma_\text{g}=0.01$).}
\end{figure*}

\begin{figure*}
\centering
\includegraphics[width=18cm,trim=0 0.7cm 0 0,clip]{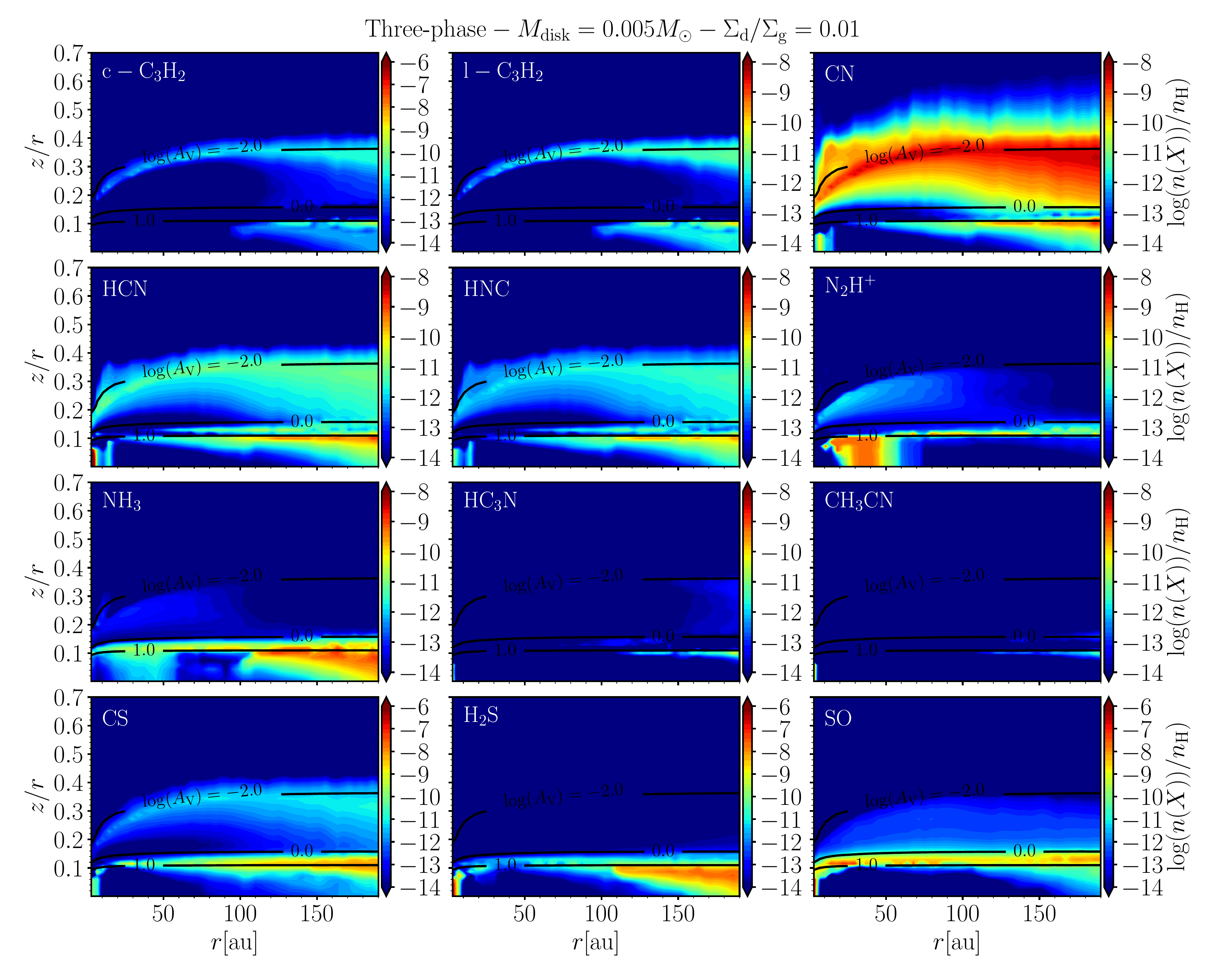}
\caption{\label{fig:ab_prof2} Continuation of Fig. \ref{fig:ab_prof1}.}
\end{figure*}

We first provide an overview of the main molecular transitions of the fiducial disk model. Gas-phase chemistry here is very similar to that in dense PDRS \citep[e.g.,][]{Tielens85, Sternberg95} and has been discussed earlier for disk conditions \citep[e.g.,][]{Aikawa99}, but is summarized here for completion and to allow quick comparison with grain surface routes to be discussed later. Figure \ref{fig:cut_r100} shows a  vertical cut at $r=50$ au of the  computed abundance profiles of some key gas-phase molecules containing hydrogen, oxygen, carbon, nitrogen and sulfur as a function of $A_V$ to the star. The H/H$_2$ transition is at $A_{V} \sim 5\times10^{-3}$ mag, the C$^+$/C/CO transition at $A_{V} \sim 10^{-2}$ mag, followed by the transition between S$^+$ and S at $A_{V} \sim 0.1$ mag and N to N$_2$ at $A_{V} \sim 0.3$ mag. Oxygen is mostly in its atomic form throughout the disk surface, while a large fraction of the atomic oxygen turns molecular at the transition between the molecular layer and the disk midplane (i.e.\  $3\lesssim A_{V} \lesssim 10$ mag). At $A_{V} \gtrsim 3$ mag most of the molecules from the gas freeze-out on the surface of grains and grain surface reactions begin to determine disk chemistry.

Figure \ref{fig:ab_prof1} and Figure \ref{fig:ab_prof2} present the computed abundances of the main species in the gas phase present in this region, and those of less abundant but commonly observed molecules in disks as a function of the disk radius $r$ and $z/r$. 

\paragraph{Oxygen chemistry} 
H$_2$O has a low gas-phase abundance, $\sim 10^{-10}$, throughout the molecular layer (Fig. \ref{fig:ab_prof1}). OH and H$_2$O formation is initiated by H$_3^+$ which is primarily formed by ionization of H$_2$ by X-rays. O$_2$ forms from OH$+$O, but reaches peak abundance deeper in the disk at $3 \lesssim A_V \lesssim 10$ mag ($z/r\lesssim 0.12$, Fig \ref{fig:cut_r100}) where grain surface chemistry dominates its formation (see \S Section \ref{sec:general_results_3}).
At $r\lesssim 20$ au where gas temperatures are higher, $\rm H_2+OH$ ($\Delta E = 1740 \text{K}$) also contributes to the formation of gas phase water. The reaction $\rm C^+ + OH$ forms CO$^+$ which reacts with H and H$_2$ to form CO and HCO$^+$ respectively; HCO$^+$ further recombines dissociatively to form CO.

\paragraph{Carbon chemistry} 
Higher in the molecular layer,  carbon chemistry is primarily initiated by the formation of CH$^+$, CH$_2^+$ and CH$_3^+$ from C$^+$ reacting with H$_2^*$ and H$_2$. CH$_2^+$ and CH$_3^+$ dissociatively recombine to form CH and CH$_2$. CH, which also forms via $\rm C + H_2^*$,  initiates carbon chain chemistry through its reaction with C$^+$ to form C$_2^+$ at $A_{V}\sim10^{-2}$ mag (see Fig. \ref{fig:cut_r100}) which then leads to the formation of C$_2$, CCH and C$_2$H$_2$.  Carbon addition can occur during these pathways to form increasingly larger molecules, such as $c-$C$_3$H$_2$/$l-$C$_3$H$_2$. Carbon chain abundances drop deeper into the molecular layer (at $A_{V}\sim 10^{-2}$ mag, see Fig. \ref{fig:cut_r100}), as CO formation reduces the availability of carbon in a reactive form.
 
 \paragraph{Nitrogen chemistry} 
CN, HCN and HNC efficiently form from carbon chains and thus closely follow their distribution (Figs. \ref{fig:cut_r100}, \ref{fig:ab_prof1} and \ref{fig:ab_prof2}). Nitrogen chemistry is initiated by the X-ray photoionization of N in the upper layers of the disk, and N$^+$ then reacts with H and H$_2$ to eventually lead to the formation of  NH, NH$_2$ and NH$_3$. CN forms via the reactions $\rm NH + C^+$ and at $A_{V}\sim10^{-2}$ mag, more eficiently from  N reacting with $\rm C_2$ and CH. CN can react with N and lead to N$_2$ (at $A_{V}\gtrsim 10^{-2}$ mag in Fig. \ref{fig:cut_r100}). Deeper in the molecular layer, nitrogen becomes fully molecular, and is also produced by $\rm N +OH$ which leads to NO, with $\rm NO+ N$ forming N$_2$.  The formation of HCN and HNC involves two main routes: (i) the neutral-neutral reaction of N with CH$_2$, and (ii) the reaction of N with CH$_3^+$ leading to HCN$^+$ which reacts with H$_2$ to form HCNH$^+$, which then dissociatively recombines to form HCN and HNC. In addition, HCN can form via the channel $\rm H+CCN$, where CCN is produced by N reacting with CCH.

 \paragraph{Sulfur chemistry} 
CS, HS and SO are the only relatively abundant sulfur bearing molecules throughout the molecular layer. In the upper layers of the disk, CS primarily forms through reaction of CH with S$^+$ leading to CS$^+$. CS$^+$ then reacts with H$_2$ to form HCS$^+$ which dissociatively recombines to CS. HS is formed through reaction of S$^+$ with H$_2$ followed by the dissociative recombination of H$_2$S$^+$. The formation of SO two main routes: (i) $\rm OH + S$ at intermediate heights and (ii) $\rm S + O_2$ when O$_2$ becomes abundant at the interface between the molecular layer and the midplane.

\begin{figure*}
\centering
\includegraphics[width=18cm,trim=0 0.7cm 0 0,clip]{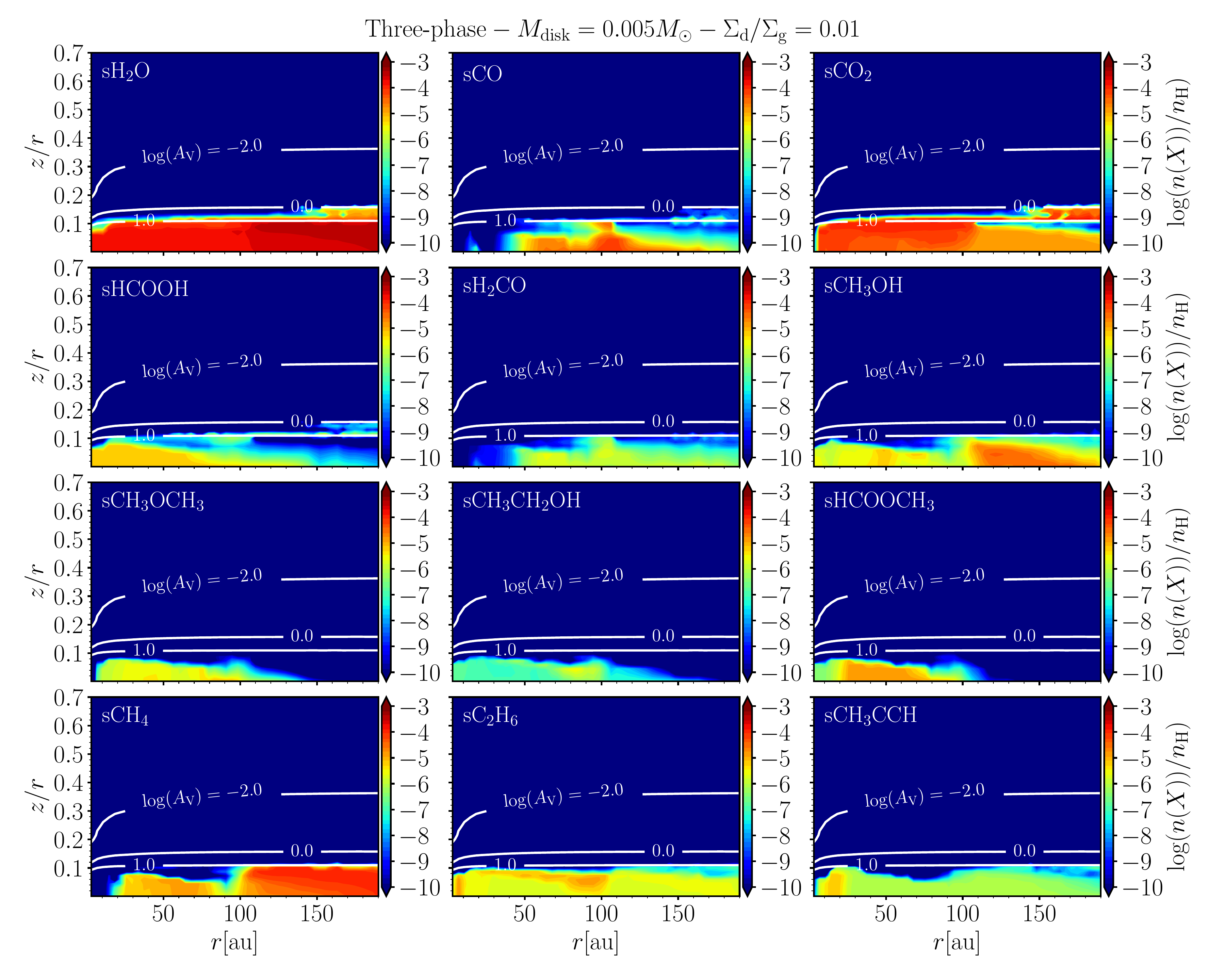}
\caption{\label{fig:ab_prof3} Computed grain surface abundances (sum of surface and mantle abundances) of a selection of molecules as a function of  disk radius $r$  and normalized height $z/r$ for the fiducial disk model ( $M_\text{disk}=0.005M_\odot$ and $\Sigma_\text{d}/\Sigma_\text{g}=0.01$).}
\end{figure*}

\begin{figure*}
\centering
\includegraphics[width=18cm,trim=0 0.7cm 0 0,clip]{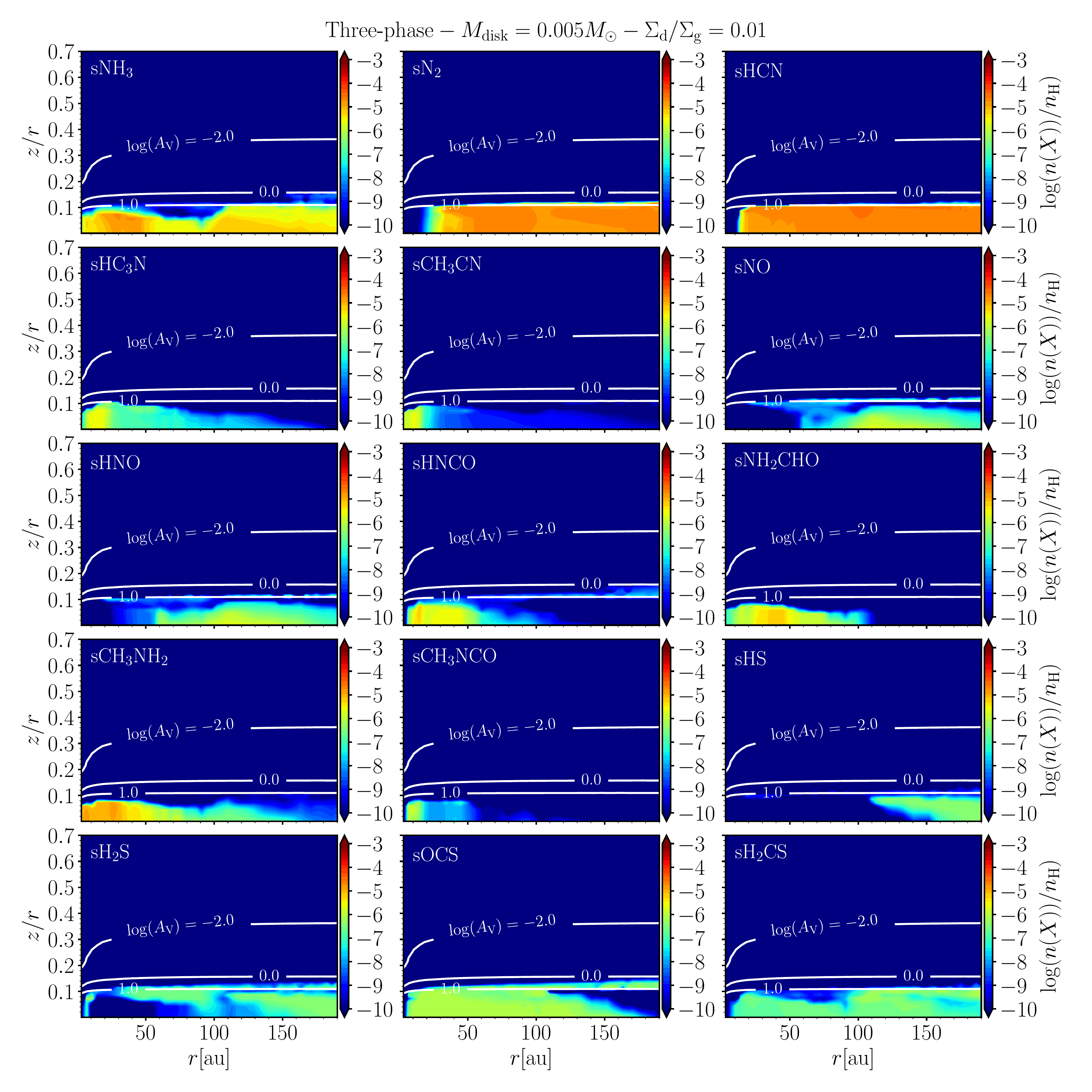}
\caption{\label{fig:ab_prof4} Continuation of Fig. \ref{fig:ab_prof3}.}
\end{figure*}

\subsubsection{Chemistry of the midplane}
\label{sec:general_results_3}
Chemistry in the midplane is characterized by ice lines or the condensation fronts of volatile species such as water, CO, CO$_2$, and N$_2$. In the inner disk midplane, the ice line location depends on the binding energy of the molecule and disk radial temperature structure; warmer dust grains are needed to release ices with higher binding energies. Water, for instance, has a relatively higher binding energy than CO (i.e.\  $E_\text{bind}({\rm H_2O}) = 5700$ K and $E_\text{bind}({\rm CO}) = 1150$ K), and therefore has an ice line at smaller radii where the temperatures are higher.  In the fiducial model, the water ice line is at $r\sim 1.6 $ au from the central star. 

However, after condensation onto grains, a molecule may not remain in the same form but could be chemically converted into other species. For instance, CO, which is the main gas phase carbon bearing molecule in disks, is thought to be desorbed to the gas-phase when $T_\text{d}\gtrsim 20$K (which represents the approximate temperature at which accretion and thermal desorption of sCO from the grains balance). But this implicitly assumes that most of the carbon is on the form of sCO in the ice. As will be discussed later, we find that sCO is efficiently channeled to sCO$_2$ (for which $E_\text{bind} = 2875$ K) in most of the disk midplane (see Fig. \ref{fig:ab_prof3}). This efficient conversion from sCO to sCO$_2$ has a strong impact on the location of the radial and vertical CO snowline, i.e.\ the radial CO snowline tends to move closer to the star as a function of time while the vertical CO snowline tends to shift higher up from the disk midplane. A notable consequence of this effect is that simple considerations based on the balance between accretion and thermal desorption may not be accurate when determining snow line locations, and chemical effects may have to be taken into account. 

Figures  \ref{fig:ab_prof3} and \ref{fig:ab_prof4} present the computed total (surface $+$ mantle) abundances of several key ice species.

\paragraph{Oxygen chemistry}
Oxygen at the grain surface is mostly in the form of sH$_2$O and sCO$_2$ (Fig. \ref{fig:ab_prof3}). sH$_2$O forms through successive hydrogenation of oxygen atoms at the surface. 
sCO is efficiently channeled into sCO$_2$ and sHOCO by reaction with sOH, through
\begin{eqnarray}
\text{sOH} + \text{sCO} &\rightarrow& \text{sCO}_2 + \text{sH}~(\Delta E = 150 \text{K}) \\
                         		    &\rightarrow& \text{sHOCO}~(\Delta E = 150 \text{K})
\end{eqnarray}
where sOH is mainly produced by photodissociation of water ice by FUV photons, including stellar, interstellar and cosmic-ray generated FUV photons. For water, dissociation by cosmic ray generated photons begins to dominate over  dissociation by stellar and interstellar FUV photon when the local $G_0$ drops below $\sim 10^{-6}$. In the fiducial model, this happens in regions located below $z/r \sim 0.08$ and for $r\lesssim 70$ au (see Fig. \ref{fig:default_disk_struc}).
sHOCO efficiently reacts with sH to form $\text{sH}_2\text{O} + \text{sCO}$,  $\text{sH}_2 + \text{sCO}_2$ and $\text{sHCOOH}$. This efficient conversion of sCO to sCO$_2$ has also been reported in several recent studies, e.g. \citet{Furuya14}, \citet{Aikawa15}, \citet{Reboussin15}, \citet{Molyarova17}, \citet{Eistrup16}, \citet{Bosman18}.

The region located between $A_{V} \sim 3$ mag and $A_{V} \sim 10$ mag (at $z/r \sim 0.12$) is characterized by photoprocessing of ices by stellar and interstellar FUV photons (see Fig. \ref{fig:default_disk_struc}) and a subsequent efficient diffusion of radicals at the grain surface ($T_\text{d}\gtrsim 15-30$ K at all radii in this vertical region where $3 \lesssim A_V \lesssim 10$ mag). Here sH$_2$O and sCO$_2$ are both efficiently photodesorbed and photodissociated. The gas phase abundance of water reaches a peak between $A_{V} \sim 3$ and $A_{V} \sim 10$ due to photodesorption of water ice (Fig. \ref{fig:ab_prof1} and Fig. \ref{fig:cut_r100}). Photodissociation of sH$_2$O and sCO$_2$ leads to the products  $\rm sOH + sH$ and $\rm sCO + sO$ respectively. 
Due to the relatively warm temperature of grains in this region, diffusion of radicals is efficient while sH has a reduced residence time at the surface and returns to the gas phase. Neutral-radical and radical-radical reactions can thus efficiently compete with hydrogenation reactions. Therefore, solid water does not efficiently reform after photodissociation and sOH mainly reacts with sCO to form sCO$_2$ and sHOCO respectively where sHOCO can then get hydrogenated to form sHCOOH. Photodesorption as well as chemical desorption of sCO$_2$ and sHCOOH then results in their enhanced gas phase abundance at $3\lesssim A_V \lesssim 10$ mag (see Fig. \ref{fig:ab_prof1}). 
sO$_2$ is also found to efficiently form in the $3\lesssim A_V \lesssim10$ mag region by 
\begin{equation}
\text{sO} + \text{sO} \rightarrow \text{sO}_2
\end{equation}
and by
\begin{eqnarray}
\text{sO} + \text{sOH} &\rightarrow& \text{sH}\text{O}_2 \\
\text{sH}\text{O}_2 + \text{sO} &\rightarrow& \text{sO}_2 + \text{sOH}
\end{eqnarray}
where sO is produced by photodissociation of sCO$_2$. sO$_2$ thermally desorbs to the gas-phase causing the high gas-phase abundance seen in Fig. \ref{fig:cut_r100}.

We note that below $A_V\sim 10$ mag (which demarcates regions with $z/r\sim 0.1$ for the fiducial model and where the interstellar UV photon flux dominate)  the opacity in the $7.5-13.6$ eV range starts to be dominated by water and CO$_2$ ice, for which the vertical column densities reach $\sim 10^{17}$ cm$^{-2}$ (the absorption cross section of CO$_2$ and $\rm H_2O$ are on the order of $\sim 10^{-18} - 10^{-17}$ cm$^{-2}$ in the range 7.5-10.3 eV for both the gas and the ice \citep{CruzDiaz14a,CruzDiaz14}. As we will see in \S \ref{sec:comp_physical_parameters}, FUV shielding by ices has an interesting implication for the column densities of gas phase species. 

As seen in Fig. \ref{fig:ab_prof3}, organic molecules like sH$_2$CO and sCH$_3$OH efficiently form on ices. These molecules form by successive hydrogenation of sCO following the sequence 
\begin{equation}
\text{sCO} \xrightarrow{\text{sH}} \text{sHCO} \xrightarrow{\text{sH}} \text{sH}_2\text{CO} \xrightarrow{\text{sH}}  \text{sCH}_2\text{OH}/\text{sCH}_3\text{O} \xrightarrow{\text{sH}}  \text{sCH}_3\text{OH}
\end{equation}
where both, $\rm sH + sCO$ and $\rm sH + sH_2CO$ reactions, show notable barriers;  $\Delta E = 2500 \text{K}$ for the reaction leading to sHCO, $\Delta E = 5400$ K for the reaction leading to $\rm sCH_2OH$ and $ \Delta E = 2200$ K for the reaction leading to $\rm sCH_3O$. However, sH can partially overcome the energy barrier via quantum tunneling, and therefore sHCO, sCH$_2$OH and sCH$_3$O are efficiently formed by these reactions\footnote{Note that hydrogen abstraction reactions can also occur and are taken into account for sHCO, $\rm sH_2CO$, $\rm sCH_2OH$, $\rm sCH_3O$ and $\rm sCH_3OH$ \citep{Chuang16}. For the H-atom abstraction reaction from $\rm H_2CO$ we use $\Delta E = 1740$ K \citep{Ruaud15}. For $\rm sCH_3OH$ we use barriers given in \citet{Garrod13}, i.e.\  $\Delta E = 4380$ K for the reaction leading to $\rm sCH_2OH$ and $\Delta E = 6640$ K for the reaction leading to $\rm sCH_3O$.}.
The presence of barriers makes the formation of methanol and its intermediates more sensitive to dust temperature than other hydrogenation reactions. The formation routes compete with the diffusion of hydrogen at the surface, and when a reaction has a barrier, its efficiency relative to sH diffusion decreases rapidly with increasing temperature,  e.g. in our model the efficiency of the reaction sH+sCO decreases by $\sim$ 4 orders of magnitude as $T_d$ changes from 10 to 20K (see Eq. D18 in Appendix \ref{sec:grain_chemistry}).
When formed at the surface (as opposed to the mantle), the products can be desorbed to the gas-phase due to chemical desorption at each step of this sequence. Desorption during formation dominates the production of gas-phase methanol in our model. 

The abundance of gas-phase methanol in the model disk is moreover strongly dependent on a continuous formation of methanol ice which then gets chemically desorbed. As seen in Fig. \ref{fig:ab_prof1}, gas phase methanol is abundant only in a thin region located beyond $r\sim 100$ au at $A_{V}\gtrsim 10$ mag, with a peak abundance of few $10^{-9}$ at $A_{V}\sim 10$ mag. 
In this region, sCO$_2$ is efficiently photodissociated by interstellar FUV photons leading to sCO and sO (similar to what was discussed earlier for the  $3 \lesssim A_{V} \lesssim 10$ mag region). Further, at larger $A_V$, grains are cold enough (i.e.\  $T_\text{d}\sim 15$ K) for hydrogenation reactions to dominate over radical diffusion (less radical mobility, larger sH residence time on grains). As a result, sO is hydrogenated to form sH$_2$O, while sCO is hydrogenated to form sCH$_3$OH. In the absence of photons, however, the production of gas-phase methanol strongly decreases because its net formation rate at the surface decreases when all the available sCO has been channeled to sCH$_3$OH. In the presence of photons, as is the case at $r\gtrsim 100$au, sCH$_3$OH is photodissociated leading to sCH$_2$OH, sCH$_3$O and sCH$_3$ \citep[sCH$_2$OH being the main outcome, see][]{Oberg09c}. sCH$_2$OH and sCH$_3$O are then hydrogenated to reform methanol, maintaining a continuous formation-destruction cycle which then allows for chemical desorption en route leading to gas-phase methanol. We note that a fraction of sCH$_2$OH and sCH$_3$O get also photodissociated leading to sCH$_2$ + sOH and to sH$_2$CO + sH respectively (relevant for carbon chemistry discussed below).
At this location, the efficient production of gas-phase methanol also contributes significantly to the production of gas-phase H$_2$CO by photodissociation of CH$_3$OH.

In the inner disk midplane ($r\lesssim100$au), grain temperatures are warm enough for an efficient diffusion of the radicals and radical-radical reactions efficiently compete with hydrogenation reactions. These regions are in general well shielded by stellar and interstellar FUV photons and cosmic-ray generated FUV dominates the production of radicals on the ices. Cosmic rays initiate the formation of more complex organics in these regions from the dissociation of the main ice species (e.g. $\rm H_2O$, $\rm CO_2$, $\rm CH_4$, $\rm CH_3OH$). In the inner disk midplane,  the radical-radical reaction
\begin{equation}
\rm sOH + sCH_3 \rightarrow sCH_3OH
\end{equation}
is found to play an important role in the synthesis of methanol ice. 
Large molecules such as $\rm sCH_3OCH_3$, sCH$_3$CH$_2$OH and $\rm sHCOOCH_3$ are also efficiently formed by radical-radical reactions (see Fig. \ref{fig:ab_prof3}). 
$\rm sCH_3OCH_3$ forms by
\begin{equation}
\rm sCH_3 + sCH_3O \rightarrow sCH_3OCH_3
\end{equation}
and
\begin{equation}
\rm sCH_3 + sHCO \rightarrow sCH_3CHO \xrightarrow{sH} sCH_3OCH_2 \xrightarrow{sH} sCH_3OCH_3.
\end{equation}
sCH$_3$CH$_2$OH forms by
\begin{eqnarray}
\rm  sCH_3 + sCH_2OH \rightarrow sCH_3CH_2OH.
\end{eqnarray}
and $\rm sHCOOCH_3$ by
\begin{eqnarray}
\rm  sHCO + sCH_3O \rightarrow sHCOOCH_3.
\end{eqnarray}

Gas-phase CO$_2$ is very abundant only at $r\lesssim 10$ au in the disk midplane, as seen in Fig. \ref{fig:ab_prof1}. 
At this location, the high temperatures allow its efficient gas-phase formation through 
\begin{equation}
{\rm OH + CO} \rightarrow {\rm CO_2 + H } ~(\Delta E = 176 \text{K}) 
\end{equation}

\paragraph{Carbon chemistry}
Most of the carbon in the disk is bound to oxygen, either as CO, CO$_2$ or in organic molecules. The main carbon-bearing ice species (without O) are sCH$_4$ and carbon chains such as $\rm sC_2H_6$ and $\rm sCH_3CCH$ (Fig. \ref{fig:ab_prof3}). sCH$_4$ is the most abundant carbon ice in the outer disk. Formation of sCH$_4$ starts with the photodissociation of sCH$_3$OH to form sCH$_2$OH, sCH$_3$O and sCH$_3$. sCH$_2$OH and sCH$_3$O can get further photodissociated to form $\rm sCH_2 + sOH$ and $\rm sH_2CO + sH$. Once formed, sCH$_2$ and sCH$_3$ are hydrogenated to form sCH$_4$ which accumulates in the outer disk midplane. 
In the inner disk midplane (i.e.\  at $r\lesssim 100$ au), the temperature of the dust is high enough for  carbon chains to form through radical-radical reactions such as
\begin{eqnarray}
{\rm sCH_2 + sCH_2} &\rightarrow& {\rm sC_2H_4}  \\
{\rm sCH_2 + sCH_3} &\rightarrow& {\rm sC_2H_5} \\
{\rm sCH_3 + sCH_3} &\rightarrow& {\rm sC_2H_6}
\end{eqnarray}
as well as
\begin{equation}
{\rm sCH_4 + sCCH} \rightarrow {\rm sC_2H_2 + sCH_3} ~(\Delta E = 250 \text{K}).
\end{equation}
$\rm sC_2H_2$, $\rm sC_2H_4$ and $\rm sC_2H_5$ can then be successively hydrogenated to from $\rm sC_2H_6$.

In a mechanism similar to that of methanol, at $A_V \gtrsim 10$ mag region in the outer disk midplane, the continuous formation/destruction cycle of sCH$_4$ and chemical desorption enhance its gas-phase abundance and subsequent gas-phase reactions and photodissociation promote the formation of carbon chains in the gas. This can be clearly seen in Fig. \ref{fig:ab_prof2} where CH$_4$, CCH, C$_2$H$_2$ show a peak in their gas phase abundances for $r\gtrsim 100$ au and around $A_V \sim 10$ mag. At this location, sCH$_4$ photodissociation leads to the formation of sCH, sCH$_2$ and sCH$_3$ (sCH$_2$ being the main outcome). These radicals then react with sH at the surface leading to the reformation of sCH$_4$. During this process species like sCH$_2$, sCH$_3$ and sCH$_4$ are chemically desorbed into the gas phase where they react to form carbon chains. C$_2$H$_4$ for instance is efficiently produced in the gas by
\begin{equation}
\rm CH + CH_4 \rightarrow \rm C_2H_4 + H
\end{equation}
where CH$_4$ originates from the grains and CH is produced by photodissociation of CH$_4$. C$_2$H$_4$ then gets photodissociated leading to C$_2$H$_2$ and CCH. In this region, c-C$_3$H$_2$ and l-C$_3$H$_2$ are also formed in the gas by successive photodissociation of CH$_3$CCH, which itself originates from grains. As seen in Fig. \ref{fig:ab_prof1}, CH$_4$ is also found to be very abundant in the gas phase at $r\lesssim20$ au where methane formed on grains is readily thermally desorbed (we use $E_\text{bind}({\rm CH_4}) = 1250$K)  and accumulates in the gas phase. At longer time (i.e. $t>10^6$ yrs), CH$_4$ gets destroyed by reaction with gas phase species such as H$_3^+$, N$_2$H$^+$ and by cosmic rays generated photons. The end product of CH$_4$ destruction is found to be primarily CO.

\paragraph{Nitrogen chemistry}
Nitrogen ice at all radii mainly consists of sNH$_3$, sN$_2$ and sHCN (Fig. \ref{fig:ab_prof4}). Molecules such as sHNCO, sHC$_3$N, sCH$_3$CN, sCH$_3$NH$_2$, sCH$_3$NCO and sNH$_2$CHO are abundant in the inner disk midplane ( $r \lesssim 100 $ au). sNH$_3$ is formed by successive hydrogenation of nitrogen atoms. During this process, sHNO is produced by
\begin{equation}
\rm sO + sNH_2 \rightarrow \rm sHNO + sH
\end{equation}
where sO is formed by photodissociation of sCO$_2$. sHNO further reacts with sH to form sNO. In the inner disk midplane, sNO is  channeled to sN$_2$ by
\begin{equation}
\rm sNO + sNH \rightarrow \rm sN_2 + sH + sO
\end{equation}
and
\begin{equation}
\rm sNO + sNH_2 \rightarrow \rm sN_2 + sH_2O
\end{equation}
which explains the lack of sNO and sHNO in the inner disk.
sHCN mainly forms through
\begin{equation}
{\rm sCN + sH} \rightarrow {\rm sHCN}.
\end{equation}
At $r\lesssim 100$ au, the reaction 
\begin{equation}
{\rm sCH + sNH} \rightarrow {\rm sHCN + sH}
\end{equation}
also contributes to the formation of sHCN.
More complex molecules form in the warmer regions of the disk. sCH$_3$NH$_2$ is found to be  formed in the inner disk by
\begin{equation}
{\rm sCH_3 + sNH_2} \rightarrow {\rm sCH_3NH_2}
\end{equation}
while sNH$_2$CHO is formed by
\begin{equation}
{\rm sHCO + sNH_2} \rightarrow {\rm sNH_2CHO}.
\end{equation}
$\rm sHNCO$ as well as $\rm CH_3NCO$ form from sOCN through
\begin{equation}
{\rm sOCN + sH} \rightarrow {\rm sHNCO}
\end{equation}
and
\begin{equation}
{\rm sOCN + sCH_3} \rightarrow {\rm sCH_3NCO}
\end{equation}
respectively. $\rm sOCN$ is itself formed by $\rm sN + sHOCO$ where sN is accreted from the gas-phase and forms by the dissociative charge transfer of $\rm N_2$ with $\rm He^+$ as well as the dissociative recombination of $\rm N_2H^+$. $\rm N_2H^+$ forms in the gas-phase by reaction of $\rm H_3^+$ and $\rm N_2$ (in this region of the disk $\rm H_3^+$ is formed through ionization of H$_2$ by cosmic-rays)

As can be seen in Fig. \ref{fig:ab_prof2}, gas-phase N$_2$H$^+$ shows a complex structure in the inner disk midplane with a peak abundance at $20 \lesssim r \lesssim 50$ au and $z/r \lesssim 0.1$. At $r<20$ au,  N$_2$H$^+$ is destroyed by CO and $\rm CH_4$. The reaction of N$_2$H$^+$ and CO leads to the formation of $\rm HCO^+$ and N$_2$ and results in a high abundance of gas-phase $\rm HCO^+$ in the disk midplane (see Fig. \ref{fig:ab_prof1}). At $20 \lesssim r \lesssim 50$ au, CO and CH$_4$ are largely depleted at the surface of the grains while N$_2$ is still present in the gas, resulting in a high abundance of  $\rm N_2H^+$ in the gas phase. At $r>50$ au, N$_2$ freezes-out on grains, and hence there is no formation of $\rm N_2H^+$. 

\begin{figure}
\centering
\includegraphics[width=7.5cm,trim=0 0.7cm 0 0,clip]{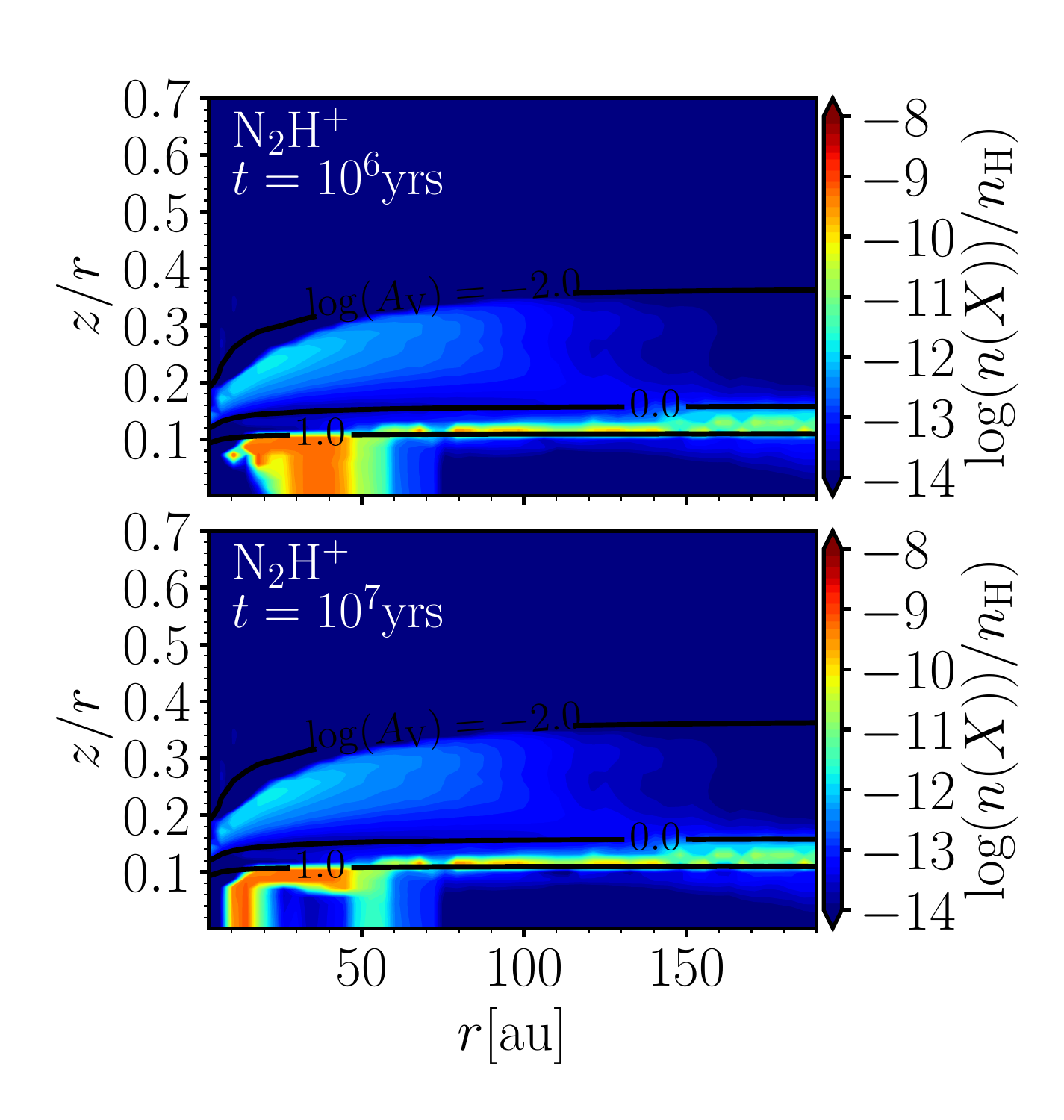}
\caption{\label{fig:n2hp} Computed abundance profile of $\rm N_2H^+$ as a function of  disk radius $r$ and normalized height $z/r$  at two different times for the fiducial disk model ($M_\text{disk}=0.005M_\odot$ and $\Sigma_\text{d}/\Sigma_\text{g}=0.01$). The abundance profile of $\rm N_2H^+$ at $t=10^6$ yrs is similar to that in Fig. \ref{fig:ab_prof2}. The $\rm N_2H^+$ region moves inward with time with the CO snowline (see \S\ref{sec:general_results_3}) }
\end{figure}

There is an additional structure  in the $\rm N_2H^+$ abundance in a narrow region located just below $A_V \sim 10$ mag and extending down to $r\sim 10$ au (Fig. \ref{fig:ab_prof2}), linked to the absence of gas phase CO (gas-phase $\rm CH_4$ has a low abundance in this region because of its efficient destruction by photodissociation). As discussed earlier photoprocessing of the ice by interstellar FUV photons is efficient in this region of the disk and sCO is quickly channeled to sCO$_2$ by reaction with sOH (which forms by photodissociation of water ice). Below this region, the FUV photon flux is not sufficient to produce enough sOH that can convert all the sCO to sCO$_2$ in the 1Myr timescale for which we have evolved the disk model. However, over longer timescales, FUV photons generated by cosmic rays can dissociate water and in fact produce a slow conversion of sCO into sCO$_2$. The dissociation timescale for water is given by 
\begin{equation}
t_\text{ph,CR} \sim 3.5\times 10^{6} \Bigg( \frac{\zeta_{\text{CR}}}{10^{-17}~ \text{s}^{-1}} \Bigg)^{-1} ~\text{yrs}
\end{equation}
and this time dependence is clearly illustrated in Fig. \ref{fig:n2hp}, which shows the change in the gas phase abundance of $\rm N_2H^+$ at  $t = 10^6$ and $t = 10^7$ yrs. The inner radial edge of $\rm N_2H^+$  is seen to have moved to $\sim 10$au, and it now traces the CO$_2$ snowline and not the CO snowline. The outer edge of $\rm N_2H^+$ also moves inward, and this is because of the conversion of N$_2$ to sNH$_3$ ice on grains; $\rm N_2$ reacts with $\rm H_3^+$ to form $\rm N_2H^+$ which then recombines with the electrons to form $\rm N_2$ and $\rm NH$. $\rm NH$ accretes at the surface of the grains and gets hydrogenated to form $\rm sNH_3$. This result was also found by \citet{Furuya14}, \citet{Aikawa15} and \citet{Eistrup16,Eistrup18}. However, \citet{Furuya14} further show that in the presence of a strong vertical turbulent mixing (i.e.\  $\alpha = 10^{-2}$ in their model), the timescales for sCO to get channeled to more refractory species is longer than the timescale for turbulent transport and that the conversion to sCO$_2$ can be reduced (the effect of vertical mixing is not taken into account in our case and left for future work).
$\rm N_2H^+$ is also found to be relatively abundant in a thin layer located at $A_V\sim 10$ mag and for $r\gtrsim 45$ au, which is also related to the  differential freeze-out of CO and N$_2$ at the surface of the grains as in the inner disk. In this layer, most of the CO is in the form of sCO$_2$ on grains, which shifts the vertical CO snowline higher up from the midplane because of the higher binding energy of sCO$_2$ relative to sCO.

As previously described for oxygen and carbon chemistry, we find that the gas-phase abundance of most of the nitrogen compounds (i.e.\  NH$_3$, CN, HCN, HNC and to a small extent HC$_3$N and CH$_3$CN) show a peak abundance at $r\gtrsim 100$ au and $A_V \sim 10$ mag. At this location, we find that the continuous formation/destruction cycle of sNH$_3$ and sHCN at the surface (i.e.\  photodestruction by the FUV photons and reformation by hydrogenation reactions) is responsible for their enhanced gas-phase abundance. When in the gas-phase, HCN gets partially photodissociated to form CN, which reacts with C$_2$H$_2$ to form HC$_3$N. CH$_3$CN is formed by dissociative recombination of CH$_3$CNH$^+$ which itself forms by reaction of HCN with CH$_3^+$. $\rm CH_3^+$ is formed by $\rm CH_2 + H_3^+$ where CH$_2$ forms from the photodissociation of CH$_4$ (which itself originates from the grains; see previous section on carbon chemistry). However, as seen in Fig. \ref{fig:ab_prof2}, neither HC$_3$N nor CH$_3$CN are particularly abundant in our model. 

\paragraph{Sulfur chemistry}

As shown in Fig. \ref{fig:ab_prof4}, sulfur is mainly in the form of sHS, sH$_2$S, sOCS and sH$_2$CS ices throughout the midplane. sHS and sH$_2$S efficiently form by hydrogenation reactions. In our model sOCS efficiently forms by
\begin{eqnarray}
{\rm sCO + sHS} &\rightarrow& {\rm sOCS + sH} ~(\Delta E = 400\text{K}) \\
&\rightarrow& {\rm sHSCO} ~(\Delta E = 400\text{K})
\end{eqnarray}
sHSCO can then be hydrogenated to form sOCS and $\rm sH_2S$. In the region around $A_V \sim 10$ mag, mainly characterized by photoprocessing of the ice, this reaction lead to an efficient desorption of these molecules at formation and explains  the enhanced abundance of gas-phase CS (where CS is formed by photodissociation of gas-phase OCS) and H$_2$S at this location seen in Figure \ref{fig:ab_prof2}.
We note that this reaction is usually not included in most chemical models because of its high endothermicity in the gas-phase. However it has been proposed to proceed with a reduced barrier on grain surfaces (see the discussion on this particular reaction in Appendix \ref{ocs_form}). 
sSO is mainly formed by
\begin{equation}
\rm sO + sHS \rightarrow sSO + sH
\end{equation}
which further reacts with sO to form sSO$_2$. 
sSO$_2$ is however efficiently destroyed by
\begin{equation}
\rm sSO_2 + sH \rightarrow sO_2+ sHS.
\end{equation}
and therefore sSO$_2$ has a relatively low abundance throughout the disk midplane.
sH$_2$CS is formed by successive hydrogenation of sCS, where sCS is formed in the gas-phase and accreted at the surface of the grains. 

In the outer disk midplane, at $r\gtrsim 100$ and $A_V \gtrsim10$ mag, sHS and sH$_2$S are  destroyed by interstellar FUV photons. sS and sHS  then get hydrogenated to reform sHS and sH$_2$S while a fraction thereof is chemically desorbed to the gas phase. Some of the sHS also reacts with sCO (produced by photodissociation of sCO$_2$) to form sOCS. The chemical desorption of sH$_2$S leads to a high abundance of gas-phase H$_2$S in this region of the disk. When in the gas-phase, a fraction of HS and H$_2$S get photodissociated to form atomic sulfur. Atomic sulfur then reacts with CH$_3$ (which is efficiently produced in this region as discussed in the section on carbon chemistry) to form CS.

As shown in Fig. \ref{fig:ab_prof2} and as discussed earlier, gas phase SO has its peak abundance in the $3 \lesssim A_{V} \lesssim10 $ mag region, which is due to the presence of abundant O$_2$ which reacts with atomic sulfur through
\begin{equation} 
\text{O}_2 + \text{S} \rightarrow \text{O} + \text{SO}.
\end{equation}

\subsection{3-phase vs. 2-phase approximation}
\label{sec:2vs3phase}
We next discuss differences between the 3-phase results presented thus far and those obtained when the same fiducial disk structure and a 2-phase chemistry approximation is used. For the 2-phase approximation, we simply assume that there is no mantle and that all the ice layers are a part of the surface. Therefore all ice layers are available for desorption, and further have the same properties for diffusion, reaction and photodissociation. Diffusion rates are computed with a diffusion-to-binding energy ratio of 0.4 (similar to the ratio used for the surface in the 3-phase model, see Appendix \ref{sec:grain_chemistry}).

\begin{figure*}
\centering
\includegraphics[width=18cm,trim=0 0.7cm 0 0,clip]{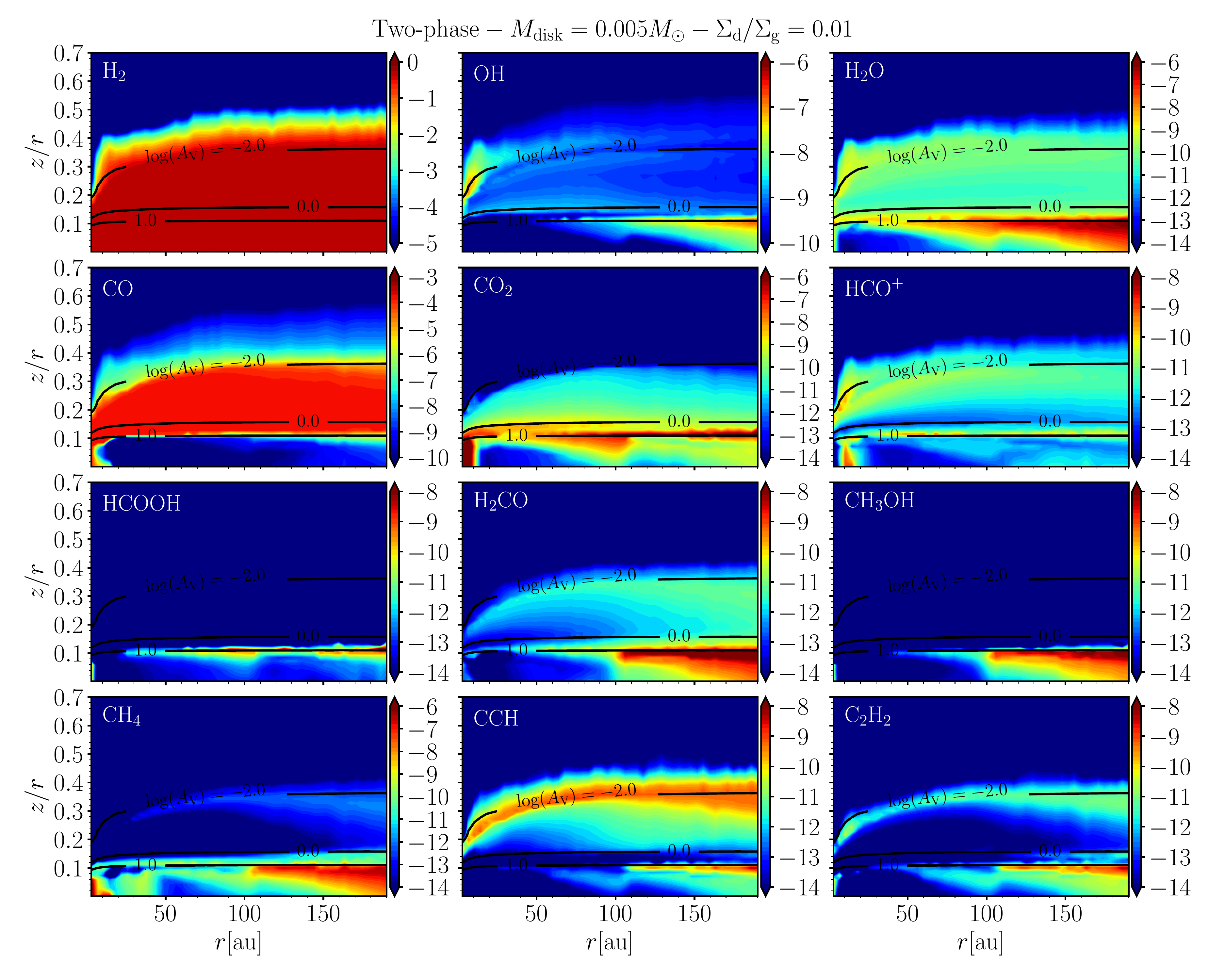}
\caption{\label{fig:ab_prof1_2phase} Gas-phase abundances of some species as a function of disk radius $r$ and normalized height $z/r$ for the Fiducial Disk model ($M_\text{disk}=0.05M_\odot$ and $\Sigma_\text{d}/\Sigma_\text{g}=0.01$) but where the 2-phase approximation has been used (similar to Fig. \ref{fig:ab_prof1} of this paper). The complete set of abundance maps for a selection gas-phase and grain surface molecules obtained with this model (4 figures) can been found in the Figure Set.}
\end{figure*}

The computed abundance maps of the main gas-phase and grain surface molecules (i.e. similar to those shown in Fig. \ref{fig:ab_prof1} to \ref{fig:ab_prof4} for the fiducial 3-phase model) can be found in online version of this article (see Fig. \ref{fig:ab_prof1_2phase}).

Fig. \ref{fig:col_dens_gas} shows a comparison of the vertical column densities, where major differences are seen between the 2-phase and 3-phase approximations for most of the observed molecules in disks. At large $r$, the vertical column densities are in general larger when the 2-phase approximation is used. The most affected molecules are CO$_2$, HCOOH, H$_2$O, HCN and HNC, for which the vertical column densities are increased by approximately two orders of magnitude at large $r$ when compared to the 3-phase model. This is mainly due to the enhanced desorption rates in the 2-phase model where all ice layers, typically on the order of a few  $10^2$ monolayers,  are available for desorption whereas only a few surface layers are available in the 3-phase model. Desorption rates and abundances are hence higher by this factor of $\sim 10^2$ as compared to the 3-phase model. Molecules such as H$_2$O, CO$_2$ and HCOOH, which are efficiently produced by photodesorption from the ice in the irradiated regions of the disk midplane are particularly affected.

For some species, such as HS and H$_2$S, we find that the two approximations lead similar results. There is again an enhanced desorption of sHS and sH$_2$S in the 2-phase model, leading to higher gas phase abundances of  HS and H$_2$S, but some of this is counterbalanced by the conversion of sHS and sH$_2$S to sH$_2$CS. 
This process is intiated by the enhanced desorption of sCH$_4$ in the 2-phase model, and the gas phase CH$_4$ produced is efficiently photodissociated to CH, CH$_2$ and CH$_3$. CH$_3$ then reacts with atomic sulfur (which is itself produced by photodissociation of HS and H$_2$S) to form CS (see discussion in the section on sulfur chemistry). CS subsequently freezes-out at the surface of the grains and is efficiently channeled to sH$_2$CS (see figures in the online version).

We also find that the predicted abundances of complex organic molecules at the surface of the grains are generally lower in the 2-phase model as compared to the 3-phase model (see figures in online version). This is because 3-phase models (or more generally multi-layer models) predict relatively higher abundances of radicals than 2-phase models \citep[see][]{Chang14,Ruaud16}, explained by the reduced mobility of hydrogen atoms in the mantle of the grains, i.e.\ hydrogenation reactions are not as efficient as in the 2-phase approximation where saturated species tend to dominate the ice composition.

\figsetstart
\figsetnum{10}
\figsettitle{Computed abundance maps of a selection gas-phase and grain surface molecules obtained for the Fiducial Disk model ($M_\text{disk}=0.05M_\odot$ and $\Sigma_\text{d}/\Sigma_\text{g}=0.01$) but where the 2-phase approximation has been used.}

\figsetgrpstart
\figsetgrpnum{10.1}
\figsetgrptitle{H$_2$ - C$_2$H$_2$ abundances}
\figsetplot{abun1_2phase.pdf}
\figsetgrpnote{Gas-phase abundances of some species as a function of disk radius $r$ and normalized height $z/r$ for the Fiducial Disk model ($M_\text{disk}=0.05M_\odot$ and $\Sigma_\text{d}/\Sigma_\text{g}=0.01$) but where the 2-phase approximation has been used.}
\figsetgrpend

\figsetgrpstart
\figsetgrpnum{10.2}
\figsetgrptitle{c-C$_2$H$_2$ - SO abundances}
\figsetplot{abun2_2phase.pdf}
\figsetgrpnote{Gas-phase abundances of some species as a function of disk radius $r$ and normalized height $z/r$ for the Fiducial Disk model ($M_\text{disk}=0.05M_\odot$ and $\Sigma_\text{d}/\Sigma_\text{g}=0.01$) but where the 2-phase approximation has been used.}
\figsetgrpend

\figsetgrpstart
\figsetgrpnum{10.3}
\figsetgrptitle{sH$_2$O - sCH$_3$CCH abundances}
\figsetplot{abun3_2phase.pdf}
\figsetgrpnote{Molecular ice abundances of some species as a function of disk radius $r$ and normalized height $z/r$ for the Fiducial Disk model ($M_\text{disk}=0.05M_\odot$ and $\Sigma_\text{d}/\Sigma_\text{g}=0.01$) but where the 2-phase approximation has been used.}
\figsetgrpend

\figsetgrpstart
\figsetgrpnum{10.4}
\figsetgrptitle{sNH$_3$ - sH$_2$CS abundances}
\figsetplot{abun4_2phase.pdf}
\figsetgrpnote{Molecular ice abundances of some species as a function of disk radius $r$ and normalized height $z/r$ for the Fiducial Disk model ($M_\text{disk}=0.05M_\odot$ and $\Sigma_\text{d}/\Sigma_\text{g}=0.01$) but where the 2-phase approximation has been used.}
\figsetgrpend

\figsetend

\subsection{Dependence of chemistry on disk physical parameters}
\label{sec:comp_physical_parameters}

Gas-grain chemistry can be expected to depend on the dust mass relative to the gas, and here we consider variations about the fiducial model. We compute three additional models: (i) a Massive Disk with 10 times higher mass ($M_{disk}=0.05$M$_\odot, \Sigma_\text{d}/\Sigma_\text{g}=0.01$), (ii) a Dusty Disk with 10 times higher dust mass ($M_{disk}=0.005$M$_\odot, \Sigma_\text{d}/\Sigma_\text{g}=0.1$), and (iii) an Evolving Disk with the same dust and gas mass as the fiducial disk model but a dust/gas ratio and dust size distribution that varies with radius. We first briefly describe the differences in physical structure of the disk in each case and then discuss the chemical differences. All the results presented here are obtained with the 3-phase model. 

\begin{figure*}
\centering
\includegraphics[width=18cm,trim=0 0.7cm 0 0,clip]{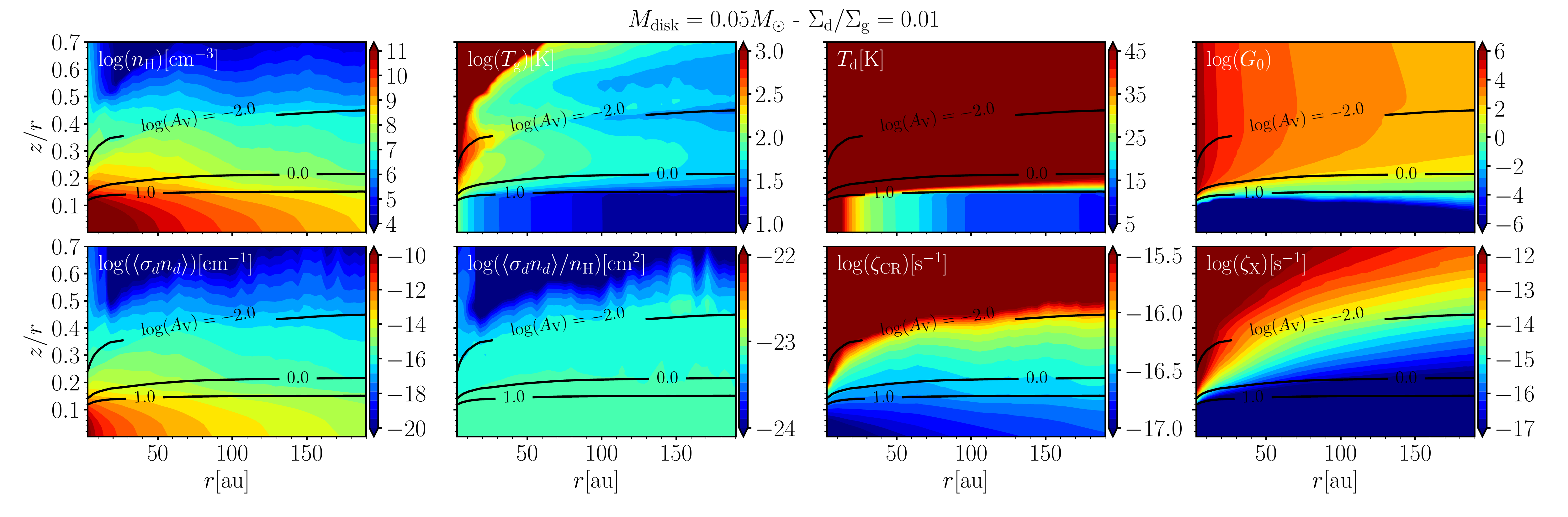}
\caption{\label{fig:disk_struc_m10} Physical structure of the Massive disk model (i.e.\ $M_\text{disk}=0.05M_\odot$ and a dust-to-gas mass ratio $\Sigma_\text{d}/\Sigma_\text{g}=0.01$) computed with the thermo-chemical model (similar to Fig. \ref{fig:default_disk_struc} of the Fiducial Disk Model). The complete set of physical structure computed for all the additional models considered in this section (3 figures) can been found in the figure set.}
\end{figure*}

\paragraph{(i) Massive Disk: $M_{disk}=0.05$M$_\odot, \Sigma_\text{d}/\Sigma_\text{g}=0.01$} This model has a higher overall gas and dust density everywhere in the disk, but the same solids/gas mass ratio as the fiducial model ($\Sigma_\text{d}/\Sigma_\text{g}=0.01$). The computed structure of the disk is presented in Fig. \ref{fig:disk_struc_m10} (i.e. similar to Fig. \ref{fig:default_disk_struc} of the Fiducial Disk Model). The computed physical structure of all the additional models considered in this section can be found in online version of this article (see Fig. \ref{fig:disk_struc_m10}). Higher gas densities in this model result in greater collisional coupling between gas and a single dust grain  (note that the grain size distribution has not been changed), and reduced Stokes numbers  which result in dust being less settled compared to the fiducial case. The disk is therefore more flared.  Increased flaring results in greater penetration of UV photons and increased heating of gas in the upper layers where the gas and dust temperatures deviate. The ionization rates by cosmic rays in this model are reduced due to a higher gas density and resulting column. This affects the chemical reset by photo-processing which is not as efficient as in our fiducial disk model; in this case the disk chemistry retains the signature of the parent cloud core chemistry after 1 Myr of the evolution. The main differences in this model as pertains to the chemistry are the flaring which impacts local optical depths ($A_V$), changes in the temperature structure, and the lower ionization in the midplane. 

\paragraph{(ii) Dusty Disk: $M_{disk}=0.005$M$_\odot, \Sigma_\text{d}/\Sigma_\text{g}=0.1$} In this model, the gas mass is identical to the fiducial case but the dust mass is 10 times higher (with the same dust mass as the massive disk model). Dust settling depends on gas density, and is the same as in the fiducial case. However, the temperature of the gas remains coupled to the dust grains at higher $z$ than the fiducial model, and these effects are especially evident at larger radii where the gas density falls. This results in a greater flaring of the disk in its outer regions compared to the fiducial disk model. Note however, that there is no appreciable change in the UV flux intercepted by the disk as its structure is nearly identical to the fiducial model in the inner regions, whereas in the outer disk, the interstellar field dominates  (see figures in online version).  The midplane cosmic ray ionization rate remains unchanged as it is a function of gas column density which is in turn unchanged from the fiducial model.

\paragraph{(ii) Evolving Disk: $M_{disk}=0.005$M$_\odot, \Sigma_\text{d}/\Sigma_\text{g}=f(r)$} Here, we consider a model disk with conditions drawn from a disk evolution model \citep{Gorti15}; we picked an epoch where the gas mass is similar to the fiducial model and where the dust/gas ratio is also similar (the dust/gas ratio ranges from 0.008 at $r=1$au to 0.012 at $r=200$au). The main difference in this case is that the gas surface density now has a self-similar profile (vs the $1/r$ power law of the fiducial model) and the dust  grain size distribution is a function of radius. The minimum and maximum grain sizes at a given $r$ are determined from a coagulation/fragmentation equilibrium approach \citep[for details see][]{Gorti15}, and the maximum grain size in particular is proportional to the local gas density; large millimeter sized grains only exist in the inner disk while the grains in the outer disk are smaller. The structure of the disk is therefore similar to the fiducial model for $r\lesssim 80$au, but deviates at larger radii  (see figures in online verion). The smaller grain sizes at large $r$ result in better coupling to gas (smaller Stokes numbers), and also a mean grain temperature that is higher in the outer disk as small grains are less efficient radiators. Both these effects increase disk flaring and hence at $r\gtrsim 80$au, the midplane temperature of the Evolving Disk is higher than that in the fiducial disk model.  
In this model, the main difference for the chemistry is that the dust grains are everywhere hotter than $\sim$25K, making radicals more mobile and competitive with hydrogenation throughout the disk. \\

\figsetstart
\figsetnum{11}
\figsettitle{Physical structure computed with the thermo-chemical model of all the additional models considered in this section.}

\figsetgrpstart
\figsetgrpnum{11.1}
\figsetgrptitle{Massive Disk model}
\figsetplot{disk_structure_massive.pdf}
\figsetgrpnote{Same as Fig. 1 of this article but for the Massive Disk model with mass $M_\text{disk}=0.05M_\odot$ and a dust-to-gas mass ratio $\Sigma_\text{d}/\Sigma_\text{g}=0.01$.}
\figsetgrpend

\figsetgrpstart
\figsetgrpnum{11.2}
\figsetgrptitle{Dusty Disk model}
\figsetplot{disk_structure_dusty.pdf}
\figsetgrpnote{Same as Fig. 1 of this article but for the Dusty Disk model with mass $M_\text{disk}=0.005M_\odot$ and a dust-to-gas mass ratio $\Sigma_\text{d}/\Sigma_\text{g}=0.1$.}
\figsetgrpend

\figsetgrpstart
\figsetgrpnum{11.3}
\figsetgrptitle{Evolving Disk model}
\figsetplot{disk_structure_epsr.pdf}
\figsetgrpnote{Same as Fig. 1 of this article but for the Evolving Disk model of mass $M_\text{disk}=0.005M_\odot$ and a dust-to-gas mass ratio $\Sigma_\text{d}/\Sigma_\text{g}=f(r)$.}
\figsetgrpend

\figsetend
 
\begin{figure*}
\centering
\includegraphics[width=18cm,trim=0 0.7cm 0 0,clip]{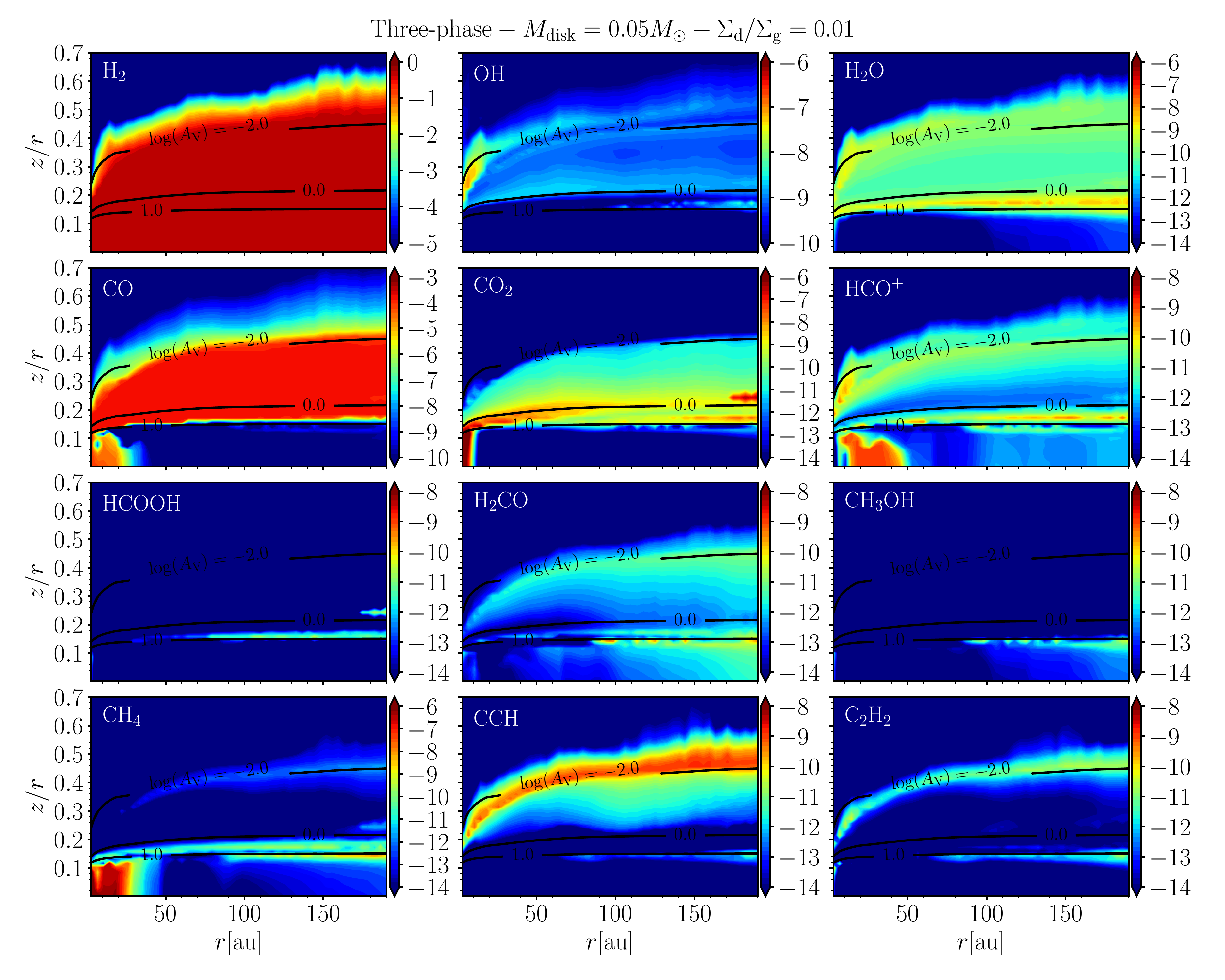}
\caption{\label{fig:ab_prof1_massive} Gas-phase abundances of some species as a function of disk radius $r$ and normalized height $z/r$ for the Massive Disk model ($M_\text{disk}=0.05M_\odot$ and $\Sigma_\text{d}/\Sigma_\text{g}=0.01$.  The complete set of abundance maps for a selection gas-phase and grain surface molecules obtained for all these models (12 figures) are in the figure set.}
\end{figure*}

\begin{figure*}
\centering
\includegraphics[width=18cm,trim=0 0.7cm 0 0,clip]{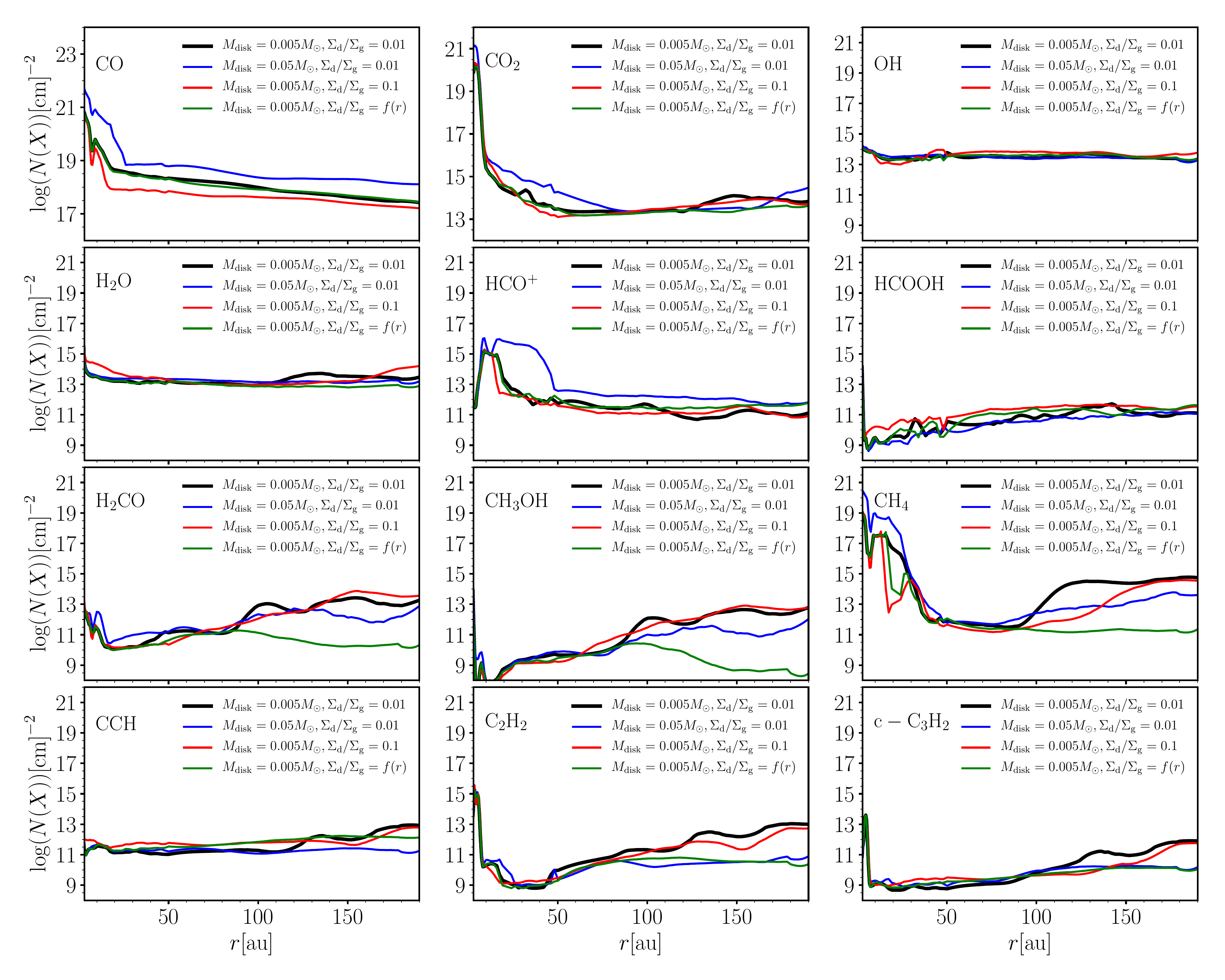}
\caption{\label{fig:disk_comp1} Computed integrated vertical column densities as a function of disk radius for a selection of gas-phase molecules. Black lines depict the fiducial model, blue lines the Massive Disk model, red lines the Dusty Disk model and green lines the Evolving Disk Model. Except for some species in the Evolving Disk model, most column densities do not show much variation between the different models.}
\end{figure*}

\begin{figure*}
\centering
\includegraphics[width=18cm,trim=0 0.7cm 0 0,clip]{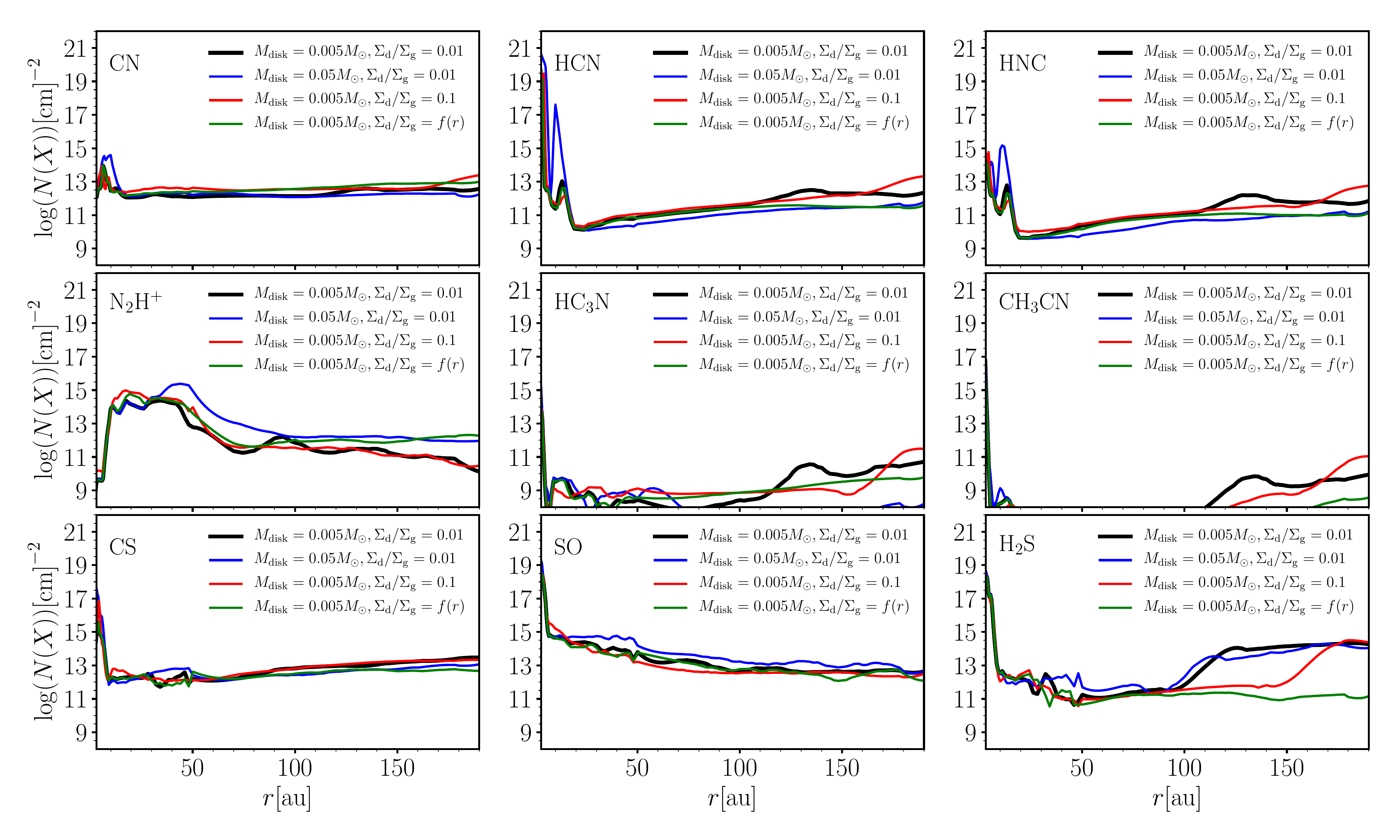}
\caption{\label{fig:disk_comp2} Same as Fig. \ref{fig:disk_comp1} but for other species.}
\end{figure*}

Figures \ref{fig:disk_comp1} and \ref{fig:disk_comp2} show the vertically integrated column densities of many gas phase molecules for all the different models. The computed abundance maps of the main gas-phase and grain surface molecules (i.e. similar to Fig. \ref{fig:ab_prof1} to \ref{fig:ab_prof4} of the Fiducial Disk Model) can be found in online version of this article (see Fig. \ref{fig:ab_prof1_massive}). The main result to be noted is that the column densities of most species appear insensitive to the disk physical parameters considered here. \\

\figsetstart
\figsetnum{12}
\figsettitle{Computed abundance maps of a selection gas-phase and grain surface molecules for all additional models considered in this section.}

\figsetgrpstart
\figsetgrpnum{12.1}
\figsetgrptitle{Massive Disk model -- H$_2$ - C$_2$H$_2$ abundances}
\figsetplot{abun1_massive.pdf}
\figsetgrpnote{Gas-phase abundances of some species as a function of disk radius $r$ and normalized height $z/r$ for the Massive Disk model ($M_\text{disk}=0.05M_\odot$ and $\Sigma_\text{d}/\Sigma_\text{g}=0.01$).}
\figsetgrpend

\figsetend

The vertical zoning of species and the column densities are both determined by the onset of formation of water and CO$_2$ ice. For $A_V\gtrsim 3$ mag, the FUV photon the photodesorption rate of water is lower than the rate of accretion of oxygen atoms at the surface \citep[see also][]{Hollenbach09}. At $3 \lesssim A_V \lesssim 10$ mag photodissociation of water ice produces sOH and warm grain temperatures increase radical mobility for a rapid conversion of sCO to sCO$_2$ via reactions with sOH. By $A_V\sim 10$ mag the integrated vertical column densities of sH$_2$O and sCO$_2$ throughout the disk reach $\sim 10^{17}$ cm$^{-2}$ and the opacity starts to be dominated by these two species (the absorption cross-section of these two molecules are on the order of $\sim 10^{-18} - 10^{-17}$ cm$^{-2}$ in the 7.5-13.6 eV range). Therefore, there are two principal disk quantities that determine the chemical layering: (i) the height $z_{ice}$  at which $A_V\sim 3$ mag  where water ice begins to form, and (ii) the column density   N$_\text{ice}$ at which water and CO$_2$ ice efficiently shield the deeper layers. The gas phase abundances of molecules as shown in Figs. \ref{fig:disk_comp1} and \ref{fig:disk_comp2} are related to the formation and desorption of ice species, and only vary as much as $z_{ice}$ and  N$_\text{ice}$ vary with disk model parameters.

The vertical column density of CO is determined by the location of the CO photodissociation front and the height $z_{ice}$ where water ice begins to form and sCO gets converted into sCO$_2$. Due to the exponential drop in density with $z$, regions closer to $z_{ice}$ dominate the contribution to the total column of CO. The gas density at $z_{ice}$ is determined by $A_V$ to the star and the dust properties of the disk to some extent influence the column density of CO because they control the flaring angle and opacity in the disk. The Massive Disk model has a surface density 10 times higher than the fiducial disk but has a larger scale height due to better coupling between gas and dust and higher flaring. Therefore, gas densities in the $ A_V \lesssim 3$ mag region  are higher only by a factor of $\sim 3$, and the total column density of CO only increases by this factor although the Massive Disk has 10 times greater mass (Fig. \ref{fig:disk_comp1}).  For the Dusty Disk model, $\Sigma_\text{g}$ is the same as the fiducial disk, but the scale height is comparatively larger because of the higher dust/gas ratio and the density in the  $A_V\lesssim 3$ mag region comparatively lower which results in a lower total column density of CO. In the Evolving Disk model, the surface densities of gas and dust are similar to the fiducial model. Although the flaring angle is higher in the $r\gtrsim 80$au region as discussed earlier, the disk structure is very similar to the fiducial disk because the contribution to the local $A_V$ is dominated by the inner disk. There are some differences in dust opacity which increases in the outer disk for the Evolving Disk model, but the CO columns are overall similar to the fiducial model.  The radial CO snowline is also located at farther radii for the Massive Disk model  (i.e.\  at $r\sim 30$ au as compared to $r\sim 20$ au in the fiducial disk model). This is due to decreased cosmic ray ionization for higher densities which leads to a decreased production of FUV photons and a slower rate of CO to CO$_2$ conversion on dust grains (see Section \ref{sec:general_results_3} on grain chemistry).

The shift in the location of CO snowline at larger radii explains the enhanced column density of $\rm HCO^+$ seen in Fig. \ref{fig:disk_comp1} at $20\lesssim r \lesssim 50$ au. Note that the shift in location of the CO snowline does not significantly impact the $\rm N_2H^+$ column density at $r<50$ au which is dominated by the $3\lesssim A_V \lesssim 10$ mag region where sCO converts to sCO$_2$ in a process dominated by stellar and interstellar FUV photons.

For molecules such as OH, H$_2$O CO$_2$, HCOOH, CN, HCN, HNC, CS and SO, the computed vertical column densities show almost similar values for all the  models and at all radii. As discussed for the fiducial disk model, all these molecules efficiently form by photoprocessing of the ice in the $3\lesssim A_V \lesssim 10$ mag region or at $A_V\sim 10$ mag for CN, HCN and HNC.  Since the formation and destruction pathways of all these molecules are driven by FUV photons, their gas phase peak abundances mainly depend on the constant column density N$_\text{ice}$ needed to attenuate the FUV photon flux and determined by the absorption cross-section of water and sCO$_2$ ($\sim 10^{-18}-10^{-17}$ cm$^{-2} $) . We note that with higher UV fields (a change in $G_0$), the location of these molecular zones will move in the disk but the column densities will not be appreciably affected as long as water ice shields itself. However, if  the dust-to-gas mass ratio is large enough ($\Sigma_\text{d} / \Sigma_\text{g}\gtrsim 1$, and 100 times larger than the ISM value) that the dust absorption cross-section per H atom ($\sigma_\text{H}\sim 5\times 10^{-22}$ cm$^{-2}$ in the ISM and $\gtrsim 10^{-24}$ in disks if grains grow to cm sizes) begins to exceed the ISM value, then dust attenuation of FUV also contributes \citep[also see][]{Hollenbach09} and the column densities of these molecules in very dusty disks is expected to be lower than the fiducial case.

For molecules such as H$_2$CO, CH$_3$OH, CH$_4$ and H$_2$S, the vertical column density show significant differences at large $r$, e.g.\ decreased by at least three orders of magnitude in the Evolving Disk model. For H$_2$CO and CH$_3$OH, the higher grain temperatures for the Evolving Disk model significantly impact their formation by successive hydrogenation of $\rm sCO$ for which the efficiency drops at $T\gtrsim 15$K (i.e.\  at $T\gtrsim 15$K, the hydrogen diffuses before reacting with $\rm sCO$ and $\rm sH_2CO$ due to the presence of barriers for these reactions (see \S \ref{sec:general_results_3}) and sCO is efficiently channeled to sCO$_2$ by reaction with sOH which is formed by photodissociation of sH$_2$O). This results in less $\rm sCH_3OH$ forming overall, and therefore lowers the column densities of $\rm H_2CO$ and $\rm CH_3OH$ for this model. This also affects the formation of sCH$_4$ which mainly forms by successive photodissociation of $\rm sCH_3OH$ followed by hydrogenation reactions, and  reduces the abundances of gas phase methane and hydrocarbons which arise due to chemical desorption of sCH$_4$. The decrease in $\rm H_2S$ column densities compared to the fiducial model can be attributed to the change in dust temperature which is higher in the midplane and favors the production of sOCS and sH$_2$CS, thereby decreasing the abundance of H$_2$S.

Finally, as discussed for the fiducial disk model, we again find that HC$_3$N and CH$_3$CN have low abundances, and resulting vertical column densities, throughout the disk and for all the disk models (i.e.\  their vertical column densities are in general lower than $10^{10}$ cm$^{-2}$ at all radii).

\section{Discussion}
\label{sec:discussion}

Molecular structure in the 3-phase gas grain chemistry models presented here depend mainly on the radial and vertical gradients in dust grain temperature and on photoprocesses in the gas phase and on ices. Locations of the snow lines of key molecules are determined by the midplane dust temperatures; e.g.\ water, CO$_2$, CO and N$_2$ condense at radii where $T_\text{d} \sim$ 100K, 55K, 20K and 18K approximately. However molecules such as CO and N$_2$ can efficiently be processed into more refractory compounds such as CO$_2$ and NH$_3$ ice and chemical effects have to be taken into account when determining snowline locations. The composition of ices changes substantially when  dust temperatures of $T_\text{d}\lesssim 15$K are attained so that radical mobility is reduced; very complex molecules can no longer form efficiently and hydrogenation dominates the chemistry. In the fiducial disk model, this temperature is reached at $r\sim 100$ au, and complex organic ices are abundant inward of this location. In the outer disk beyond 100 au, water, CO, CO$_2$, N$_2$ and HCN ices dominate and over time CO gets hydrogenated to methanol which in turn slowly converts to methane ice on $\sim$ Myr timescales. Vertically, many species like CN, C$_2$H show a distinct two-layer structure originating from heights near the CO photodissociation layer and deeper in the disk where CO disappears from the gas phase. Since the gas densities are much higher at low $z$, the column densities of gas phase species are dominated by the lower layer.  Gas phase columns of commonly detected species are remarkably constant even when gas densities and dust-to-gas mass ratios in the disk are changed by an order of magnitude. We find that this is because their abundances are determined largely by photoprocessing of ices in the disk midplane and subsequent desorption of photoproducts. The onset of freezing is determined by the formation of water ice at A$_V \sim 3-10$. From this point, water and CO$_2$ ice start to  efficiently shield deeper regions from UV photons and the constant column density of gas needed for shielding implies that the column densities of gas phase species in this layer does not appreciably change with disk conditions.

\subsection{Comparison with previous models}
\label{sec:compmodel}

Our treatment of both dust temperature and photo-processes is different from previous  work that has considered gas-grain chemistry (see Appendices A and C). The models most similar to those presented here are by \citet{Semenov11}, \citet{Walsh12,Walsh14} and \citet{Wakelam19}. Of these, the former two are 2-phase models while \citet{Wakelam19} consider a 3-phase model. \citet{Semenov11} present a detailed study of the effects of radial and mixing on gas-grain chemistry, but also present laminar models.  Although we consider dust settling in a similar parametric $\alpha$ disk model, we do not consider mixing and therefore compare our results with the laminar model of \citet{Semenov11}. \citet{Walsh12,Walsh14} in the other hand do not consider mixing, but use an extended full gas-grain network on a disk structure that was obtained from the thermo-chemical model of \citet{Nomura05}. We note that  the underlying disk structure (density, temperature and radiation field) is quite different in each of these works from ours. However, based on our limited investigation of disk parameters, we expect disk chemical structure to be relatively insensitive to disk conditions and that comparisons between models can nevertheless be made.

As already mentioned, both \citet{Semenov11} and \citet{Walsh12,Walsh14} use the 2-phase approximation and we can attribute some of our differences with these prior results to the use of the 2-phase approximation. As discussed earlier, this results in an increase in the abundances of most species that are a result of grain surface chemistry. The consequence is that column densities of H$_2$O, CO$_2$, HCOOH, HCN, HNC are in general all lower for our 3-phase model than the laminar model of \citet{Semenov11} and \citet{Walsh12} by two orders of magnitude. In our models, formaldehyde is produced along with methanol on similar pathways, and its abundance peaks in the outer disk (i.e.\  at $r\gtrsim 100$ au in most of our models), presenting a ring-like structure. This result agrees with that presented in \citet{Walsh14} (even though the ring structure is not as marked as in our case) but contradicts \citet{Semenov11} who predict that $\rm H_2CO$ is present in a narrow region extending uniformly throughout the disk midplane with a nearly constant abundance of few $10^{-9}$. As in our models, \citet{Walsh14} predict that molecules such as $\rm sCH_3CHO$, $\rm sNH_2CHO$ or  $\rm sCH_3NH_2$ efficiently form in the ice when the dust temperature is high enough. However, \citet{Walsh14} also predict that these molecules could be present in disks with a gas-phase abundance on the order of $10^{-10}-10^{-9}$ relative to $n_\text{H}$. In \citet{Walsh14}, these molecules form at the surface of the grains throughout the disk midplane and are efficiently photodesorbed to the gas-phase in a narrow region within the molecular layer. In our  models, the production of the most complex organic molecules is restricted to the inner disk (i.e.\  $r\lesssim 100$ au for the fiducial run), where the dust temperature is high enough for efficient diffusion of the radicals at the surface. In this region, the FUV flux is usually too low for photodesorption\footnote{We use a  photodesorption yield 30 times lower than the \citet{Walsh14} value of $3\times 10^{-3}$, as recent experiments indicate that a yield of $\sim 10^{-3}$ for complex organic molecules could be an overestimate \citep[e.g.\ see][for methanol]{Bertin16,CruzDiaz16}.}  to be efficient and the calculated gas-phase abundances of all these large complex organic molecules are in general very low throughout the disk (i.e.\  less than few $10^{-14}$ with respect to $n_\text{H}$). 

\citet{Wakelam19} recently presented a parametric study of disk physical properties using a gas grain model which our network is also based upon, although we have  substantially updated the network as described in the appendices. 
They find that the initial chemical composition of the disk, dust settling and grain growth all have strong effects on the disk chemical structure. Our results do not show a dependence on the initial chemical composition, except in the inner disk regions where cosmic ray ionization is low \citep[similar to][]{Eistrup16}. We do find that dust properties affect the onset of water ice freezing and to that extent computed abundances, but we find only a weak logarithmic sensitivity to $A_V$; as in \citet{Hollenbach09}. 
Although the underlying disk chemical networks are similar, almost all the physical processes that the disk chemistry is sensitive to -- UV photochemistry, X-ray photochemistry, grain size range and growth due to condensation, coupled gas/dust physical structure -- are in fact very different between the two models, therefore producing very different results.

\subsection{Comparison with observations}
\label{sec:comp}
Most of the molecules observed  in disks around young stars  are simple compounds such as CO, $\rm HCO^+$, CN, HCN, HNC, N$_2$H$^+$, CCH, CS and SO. These have been mainly detected at (sub)-millimeter wavelengths and therefore trace the cold outer regions of protoplanetary disks where grain surface chemistry could  play a role. We use the values compiled by \citet{Agundez18} who provide an updated and comprehensive summary of most of the molecules observed in T Tauri and Herbig Ae/Be disks (see their Table 3). The predicted column densities for molecules such as HCO$^+$, HCN and HNC lie in general in the range of observed values. The ratio $N({\rm HNC})/N({\rm HCN})$ is of the order of $\sim 0.1 - 0.4$, which is in good agreement with the ratio derived in the disk orbiting TW Hya, for which $N({\rm HNC})/N({\rm HCN}) \sim 0.1-0.2$ \citep{Graninger15}. For CN, the predicted vertical columns are at the low end of those inferred from observations, $N({\rm CN})\sim 10^{13}-10^{14}$ cm$^{-2}$ \citep[see Table 3 of][]{Agundez18}. 
For CCH we find that the three models predict vertical column densities a factor of $\sim 10$ lower than the typical range derived from observations which suggest $N({\rm CCH})\sim 4\times10^{12} - 10^{14}$ cm$^{-2}$ at millimeter wavelengths. $\rm c-C_3H_2$ was also detected in TW Hya by \citet{Bergin16} but the inferred column densities are not provided.

$\rm N_2H^+$ has been observed in number of disks and believed to trace the CO snowline \citep[e.g.][]{Qi13}. As discussed earlier, we find that sCO is efficiently channeled to sCO$_2$ in most of the disk midplane which impacts the location of the CO snowline. The radial CO snowline presents a complex structure that varies with time, affecting the radial distribution of $\rm N_2H^+$ in the disk. Our models predict a column density of $\rm N_2H^+$ of order $10^{14}-10^{15}$ cm$^{-2}$ at at $10<r<50$ au. We also find that $\rm N_2H^+$ is present well beyond the $\rm N_2$ snowline with column densities on the order of $10^{11}-10^{13}$ cm$^{-2}$, originating near the vertical water condensation front. This extended distribution is in good agreement with the extended emission observed by \citet{Qi13} in the disk around TW Hya. The predicted spatial distribution; i.e.\  near the water condensation front, is also supported by the recent derivation of the temperature at which the observed $\rm N_2H^+$ lines emit in this object \citep{Schwarz19}. 

For sulfur bearing species the predicted column densities of CS and SO lie in the observed range of column densities for which  $N({\rm CS})$ and $N({\rm SO})$ are on the order of $\sim 10^{12} -10^{13}$ cm$^{-2}$. $\rm H_2S$ has recently been detected in the disk around GG Tau A with $N({\rm H_2S}) = 1.3\times 10^{12}$ cm$^{-2}$ \citep{Phuong18} which lie in the range of predicted column densities (even though we observe strong variations in the predicted column densities with disk physical parameters).

Cold water has also been observed by {\it Herschel}/HIFI in the disk orbiting TW Hya \citep{Hogerheijde11}, DG Tau \citep{Podio13} and HD 100546 \citep{Du17} and thought to originate from the outer part of these disks. Upper limits have also been derived in $\sim 10$ other objects \citep{Du17}. For disks with positive detections, the range of inferred abundances is relatively low, i.e. \ $n({\rm H_2O})/n_\text{H}\sim 5\times10^{-8} - 2\times10^{-7}$,  as compared to that predicted by previous 2-phase chemical models for which $n({\rm H_2O})/n_\text{H}\sim 10^{-7} - 10^{-5}$ in the outer parts \citep[e.g.][]{Walsh12,Semenov11}, and more in agreement with our 3-phase model results. The predicted column densities of water in all our 3-phase disk models are on the order of $N({\rm H_2O})\sim 10^{13}$ cm$^{-2}$ in the outer parts, with peak abundances on the order of $\sim 10^{-8}$ (see Fig. \ref{fig:ab_prof1} for the fiducial disk model) which is in line with the observations in these  disks.

Only a handful of molecules composed by four or more atoms have been observed at (sub)-millimeter wavelengths in disks; $\rm H_2CO$, HCOOH, HC$_3$N, $\rm CH_3OH$ and $\rm CH_3CN$. Formaldehyde has been observed in number of disks around T Tauri stars. The predicted column densities of $\rm H_2CO$ lie approximately in the range of values derived from observations from which $N({\rm H_2CO})\sim 10^{12} - 10^{13}$ cm$^{-2}$. HCOOH has also been recently detected in TW Hya with $N({\rm HCOOH}) \sim 2-4\times 10^{12}$ cm$^{-2}$ \citep{Favre18}, which is approximately an order of magnitude higher than our predictions. $\rm CH_3OH$ has recently been observed in TW Hya with an inferred column density of $N({\rm CH_3OH})\sim 3-6\times10^{12}$ cm$^{-2}$ \citep{Walsh16}, which is in good agreement with our modeling results. We also note that for H$_2$CO and CH$_3$OH our model predicts ring like strutures as they tend to form more efficiently on cold dust grains; i.e. \ with $T_\text{d}\lesssim 15$ K (see \S \ref{sec:general_results_3}). HC$_3$N and CH$_3$CN were extensively observed recently in disks around T Tauri and Herbig Ae/Be stars \citep{Bergner18}. From the observations obtained in the disk around V4046 Sgr, the authors derived column densities in between $10^{12}$ and few $10^{13}$  cm$^{-2}$ for both molecules which are at least three orders of magnitude higher than the values we predict with our 3-phase models.

\subsection*{}
To conclude, the models presented here are overall in reasonable agreement with observations, although we note that  protoplanetary disks often exhibit considerable diversity in their molecular content. We find that the dust temperature (and hence possibly dust evolution) is one of the few disk parameters that affects molecular gas phase abundances, and therefore that chemistry in the disk likely evolves as dust collects to form planets.  Although a typical T Tauri disk was considered here (with a few additional models) we expect that 
the general chemical structure calculated should hold in general for all disks, with gas phase abundances of molecules peaking in distinct regions near the CO dissociation front and near the vertical water snowline.  
We also note that the chemical models make use of several poorly known input quantities (e.g., reaction rates, binding energies, molecular opacities). In particular, FUV molecular opacities of condensed-phase species need to be better determined by laboratory experiments/theory. However, since the physical structure and chemistry is very interlinked, we can hope to make progress by simultaneous multiwavelength modeling  of disk line emission and dust continuum data to disentangle the many unknown physical and chemical parameters. Such comprehensive modeling of individual disks using the framework presented here will be the subject of future work.

\section{Conclusion}
\label{sec:conclusions}

The effects of gas-grain chemistry in protoplanetary disks were studied using a new framework that solves for the disk physical structure and dust/gas coupling using a thermochemical model, and detailed gas phase and ice chemistry using a 3-phase gas-grain network. We considered a fiducial model with stellar and disk parameters typical for T Tauri stars, and three other models where we varied the disk physical parameters (gas mass, dust mass, dust distribution).  We show that the disk chemical composition depends mainly on the radial and vertical gradients of the dust temperature and on photoprocesses in the gas-phase and on ices. Our main results can be summarized as follows:
\begin{enumerate}
\item  The disk interior can be divided into three chemically distinct regions: (1) The interface between the molecular layer and the vertical water snowline (i.e.\ located at $3\lesssim A_V \lesssim 10$ mag) where grain surface chemistry is driven by photoprocessing of ices and efficient radical diffusion, (2) the  outer disk midplane (i.e.\  at $r\gtrsim 100$ au in the fiducial model) where the dust temperature is $\lesssim 15$ K and the FUV flux is still substantial, and (3) the inner disk midplane (i.e.\  at $r\lesssim 100$ au in the fiducial model) where the dust is warm enough for radical mobility and allows the development of a complex chemistry.

\item At the interface between the molecular layer and the midplane (at $3\lesssim A_V \lesssim 10$ mag), photoprocessing of ices has a significant impact on the gas-phase abundance of several molecules. Here, water ice photodissociation and photodesorption by stellar and interstellar FUV photons is important. Water ice photodissociation leads to the production of sOH which rapidly reacts with sCO to convert most carbon into sCO$_2$. Some sHCOOH also forms and these molecules are desorbed, increasing their gas phase abundance in this region.

\item In the outer disk midplane, low dust temperatures ($T_d<15$K) result in hydrogenation dominating disk chemistry. The upper layers of the midplane (at $ A_V \sim 10$ mag in the fiducial model) are exposed to sufficient interstellar FUV photons to set a continuous formation/destruction cycle for species such as $\rm sCH_4$, $\rm sH_2CO$, $\rm sCH_3OH$, $\rm sNH_3$, $\rm sHCN$ and $\rm sH_2S$. Photons dissociate these molecules, while hydrogenation continually replenishes them. Chemical desorption releases a small fraction of these species (and reaction intermediates) to the gas phase where chemical reactions lead to an efficient gas-phase formation of species such as CCH, $\rm C_2H_2$, $\rm C_3H_2$ and $\rm CN$ in the upper layers of the disk midplane.

\item In the inner disk midplane, ($r\lesssim 100$ au in the fiducial model) the dust temperature is high enough for radicals to efficiently diffuse at the grain surface and form large, complex organic molecules such as $\rm sCH_3OCH_3$, $\rm sCH_3CH_2OH$, $\rm sHCOOCH_3$, $\rm sNH_2CHO$,  $\rm sCH_3NH_2$ and $\rm sCH_3NCO$. Although they are abundant on ices, their predicted gas-phase abundance is negligible, in general lower than few $10^{-14}$ with respect to $n_\text{H}$.

\item The conversion of sCO to sCO$_2$ impacts the location of the CO snowline which presents a complex structure that varies with time. sCO$_2$ is the dominant form of carbon in regions where water ice forms and can also be photodissociated ($r\gtrsim 10$ au and $A_V\sim 3-10$ mag in the fiducial model).
When FUV photon flux is too low, the production of sOH by cosmic rays is the rate limiting step and sCO converts to sCO$_2$ on long timescales ($\sim$ Myr) and the radial CO snowline slowly moves inward. The snow line location may thus be determined by chemistry, and considerations based on the balance between accretion and thermal desorption may not be reliable. 

\item Most commonly observed molecues are found to arise from two distinct chemical zones.  One is in the PDR region near the CO photodissociation front where there is rich gas-phase carbon chemistry, and the other is near the water condensation front where there is rich gas-grain chemistry with some products being desorbed into the gas phase. Some species, in particular, CN and C$_2$H, are abundant in both zones and show a two-layered structure. 

\item Due to the above chemical zoning, the vertically integrated column densities of many commonly observed trace gas phase species are found to be relatively insensitive to the gas mass or dust mass in the disk and are set by self-shielding and dust attenuation columns of UV photons (both in the PDR region and near the water ice region). The abundances of some molecules such H$_2$CO, CH$_3$OH, CH$_4$ and H$_2$S, which are affected by the efficiency of H diffusion and radical mobility, depend on the dust temperature in the disk. 

\item We find that calculated column densities of gas phase species from the models are in the range of what is commonly inferred from disk observations. Using the three phase approach (versus the commonly used 2-phase gas grain modeling where there is no distinction between ice surface and ice mantle) results in lower gas phase abundances of molecules that result from gas grain chemistry, notably, H$_2$O, CO$_2$, HCOOH, H$_2$CO, CH$_3$OH, CN, HCN, HNC, HC$_3$N and CH$_3$CN. We find that complex N-bearing species (i.e.\ HC$_3$N and CH$_3$CN) are under-abundant in the models presented here, and will be the subject of future investigations where we model line emission from a few observationally well-studied disks. 

\end{enumerate}

\acknowledgments
We are thankful to the anonymous referee for a thorough and constructive report which greatly improved the quality of this manuscript. The authors acknowledge support by the National Aeronautics and Space Administration through the NASA Astrobiology Institute under Cooperative Agreement Notice NNH13ZDA017C issued through the Science Mission Directorate.
M.~Ruaud's research was supported by an appointment to the NASA Postdoctoral Program at NASA Ames Research Center, administered by Universities Space Research Association under contract with NASA.

\appendix

\section{Thermochemical model details}
\label{sec:thermochem}
The thermochemical disk model used here is as described in earlier work \citep{Gorti04, Gorti08, Hollenbach09b, Gorti15}, with two modifications. The cosmic ray ionization rate has been updated to follow the prescription of \citet{Padovani18} and described in Appendix \ref{sec:cosmicray}, and the photocross-sections of atoms and molecules which were previously from \citet{Huebner92}, have been replaced with those from the LAMDA database. We summarize the main aspects of the model below. 

The disk structure is determined from the prescribed surface density distributions of gas and dust (the dust/gas ratio $\Sigma_\text{d}/\Sigma_\text{g}$). The density and temperature structure is calculated by imposing vertical hydrostatic pressure equilibrium; thermal balance of dust and gas and chemistry are solved iteratively until convergence. The effects of both FUV and X-rays photons on disk heating and chemistry are included, and gas and dust temperatures are determined separately. We include heating, photoionization and photodissociation of gas species due to EUV, FUV and X-rays (more details in Appendix \ref{sec:gas_chemistry}), thermal energy exchange by gas-dust collisions, grain photoelectric heating of gas by FUV incident on PAHs and very small grains, and heating due to exothermic chemical reactions. Cooling of gas is by line emission from atoms, ions and molecules, with steady state abundances obtained using a reduced chemical network of $\sim$ 90 species of H, He, C, O, Ne, S, Mg, Fe, Si, and Ar and $\sim$ 500 reactions. The network is based on that of \citet{Kaufman99,Kaufman06} with updates as described in \citet{Gorti04}, \citet{Gorti08}, \citet{Hollenbach09b} and \citet{Gorti11}. No grain surface chemistry is considered in this chemical network which is constructed for accurate computation of gas temperature, this is mainly because in the regions where freeze-out is important, gas and dust temperatures are typically well coupled.
Dust radiative transfer is implemented in a 1+1D model as described in \citet{Dullemond01}, but with a range of grain sizes using the absorption coefficients (Q's) calculated from refractive indices for silicates and carbon grains obtained from the Jena database\footnote{\url{https://www.astro.uni-jena.de/Laboratory/Database/jpdoc/f-dbase.html}} \citep{Henning99}.
Gas radiative transfer is an nLTE treatment using an escape probability formalism, and takes into account the dust radiation background \citep[as in][]{Hollenbach91}. Dust settling for each grain size bin is calculated from a maximum attainable scale height based on coupling to gas \citep[e.g.][]{Estrada16}, and the vertical distribution is iteratively solved to conserve the local dust surface density. Dust grain size evolution (growth/fragmentation equilibrium) is calculated as described in \citet{Gorti15} and drawn from the framework of \citet{Birnstiel10,Birnstiel12}. For further details, see \citet{Gorti04}, \citet{Gorti08} \citet{Gorti11} and \citet{Gorti15}.

\section{Cosmic rays ionisation}
\label{sec:cosmicray}
For the ionisation by cosmic rays we follow \citet{Padovani18}. They model the propagation of cosmic rays particles through a semi-infinite medium and provide a simple analytical expression for the cosmic ray ionization rate as a function of the gas column density.  This expression depends on the assumed cosmic ray spectrum, for which they consider two cases, a low energy cosmic rays spectrum and a high energy cosmic rays spectrum. We use the expression obtained for the low energy cosmic rays spectrum which fit the Voyager 1 data, i.e.\  denoted as model $\mathcal{L}$ in \citet{Padovani18}. This expression is given by
\begin{equation}
\label{eq:zeta_padovani}
\log \zeta_\mathrm{CR} = \sum_{k \geq 0} c_k \log^k N(\text{H}_2)
\end{equation}
where the coefficients $c_k$ are given in Table F.1 of \citet{Padovani18}. 


\section{Gas phase chemistry}
\label{sec:gas_chemistry}

\subsection{Gas-phase photoprocesses}
\label{sec:photoprocesses}

For the attenuation of FUV by atoms and molecules we follow the method described in \citet{Gorti04} and split the UV-visible region into nine energy bins from 0.74 to 13.6 eV. The nine intervals were chosen to correspond to dominant photoabsorption thresholds and include 0.74-2.6 eV, 2.6-3.5 eV, 3.5-4.3 eV, 4.3-5.12 eV, 5.12-7.5 eV, 7.5-10.0 eV, 10.0-10.36 eV, 10.36-11.26 eV and finally 11.26-13.6 eV. FUV photon transport is done using a 1+1D model.
When data are available, photoabsorption, photodissociation and photoionisation cross sections are integrated over each energy bin to get the weighted cross sections for each species in each bin $n$
\begin{equation}
    \langle \sigma_{i,n} \rangle = \frac{\int_{\lambda_n} \sigma_i(\lambda) \lambda I(\lambda) d\lambda}{\int_{\lambda_n} \lambda I(\lambda)  d\lambda}
\end{equation}
where $I(\lambda,r,z)$ is the mean intensity of the radiation as a function of $\lambda$ at a position $(r,z)$ in the disk. These normalized cross sections are subsequently used to compute the absorption of the FUV radiation flux by the atoms and molecules of the gas in each bin $\tau_n(r,z)$ by
\begin{equation}
    \tau_n = \sum_i \langle \sigma_{i,n}^\text{abs} \rangle N_i
\end{equation}
where $N_i(r,z)$ is the column density of the species $i$ (in the vertical and radial directions) and to compute the photodissociation and photoionisation rate coefficients by
\begin{equation} \label{eq:gas_photo_sig}
    k_{\text{ph}}(i) = \sum_{n} \langle \sigma_{i,n} \rangle F_{\text{FUV},n} e^{-(\beta A_{V} +\tau_{n})}
\end{equation}
where the term $e^{-\beta A_V}$ accounts for the continuum attenuation by the dust and  $F_{\text{FUV},n}(r,z)$ is the integrated FUV flux in each bin $n$
\begin{equation} 
    F_{\text{FUV},n} = \frac{1}{hc}\int_{\lambda_n} \lambda I(\lambda) d\lambda
\end{equation}
in photon s$^{-1}$ cm$^{-2}$. All these quantities are calculated at each location of the disk and for each component of the FUV radiation flux, i.e.\  the stellar component and the contribution coming from the ISRF in the up and down directions. 

The cross sections are taken from the Leiden database for photodissociation and photoionisation of atoms and molecules of astrophysical interest \citep{Heays17}. When data are not available we use the approximated parametric expression
\begin{equation} \label{eq:gas_photo}
     k_\text{ph}(i) = G_0 \alpha e^{-\beta A_V}
\end{equation}
where $G_0$ is the integrated FUV flux with respect to that of the ISRF in units of the Draine field in the 6-13.6 eV range and $\alpha$ is the unattenuated photorate taken from the KIDA database\footnote{\href{http://kida.obs.u-bordeaux1.fr}{http://kida.obs.u-bordeaux1.fr}}.

Effect of self-shielding on the photodissociation rates is taken into account for H$_2$, CO and N$_2$. H$_2$ self-shielding is treated following \citet{Tielens85}. For CO and N$_2$ we use the shielding functions tabulated in \citet{Visser09} and \citet{Li13} which includes the effect of self-shielding as well as the shielding provided by H and H$_2$ (mutual shielding).

\subsection{X-ray photoreactions}

Gas-phase X-ray chemistry has been added to the gas-grain chemical model following \citet{Gorti04}. The absorption cross section of a species $i$ for ionization is approximated by
\begin{equation}
    \sigma_i(E) = \sigma_{0,i}\Bigg( \frac{E}{1~\text{keV}} \Bigg)^\alpha
\end{equation}
where values for $\sigma_{0,i}$ for each atoms can been found in Table 8 of \citet{Gorti04}. The rate of photoionization is given by
\begin{equation}
    k_\text{X}(i) = 6.25\times 10^8 \int_{0.5}^{10} \sigma_i(E) \frac{F(E)}{E} e^{-\tau_\text{X}(E)} dE 
\end{equation}
where $F(E)$ is the X-ray flux in units of erg s$^{-1}$ keV$^{-1}$, $E$ the energy in keV and $\tau_\text{X}$ the X-ray optical depth given by $\tau_\text{X} = \sigma(E) N_\text{H} $, with $\sigma(E)$ the total X-ray photoabsorption cross section. For hydrogen and helium, secondary ionizations are taken into account \citep[see][for more details]{Gorti04} and the rate of photoionization are given by
\begin{eqnarray}
    k_\text{X}(\text{H}) &=& 6.25\times 10^8 \int_{0.5}^{10} \sigma_\text{H}(E) \frac{F(E)}{E} e^{-\tau_\text{X}(E)} (1 + 25.7E)dE \\
    k_\text{X}(\text{He}) &=& 6.25\times 10^8 \int_{0.5}^{10} \sigma_\text{He}(E) \frac{F(E)}{E} e^{-\tau_\text{X}(E)} (1 + 2.0E)dE
\end{eqnarray}
We note that secondary ionization of all other species is ignored. As in \citet{Gorti04}, we assume that photoionization rates of molecules are given as the sum of the photoionization rates of their constituting atoms, e.g. $k_\text{X}(\text{AB}) = k_\text{X}(\text{A}) + k_\text{X}(\text{B})$.  
We note that we consider only gas phase X-ray reactions at present and do not consider X-ray reactions of ice species.
\subsection{Reactions with PAHs}

For reactions with PAHs we follow \citet{Draine87} and \citet{Weingartner01}. From \citet{Weingartner01}, when an ion $X^i$ collide with a grain of radius $a$ and charge $Ze$, the ion removal (or neutralization) rate is given by
\begin{equation}
\frac{dn(X^i)}{dt} = - \int dn_d \sum_Z f(a,Z)J_i(Z)\theta [\text{IP}(X^{i-1}) - \text{IP}(a,Z)]
\end{equation}
where $f(a,Z)$ is the fraction of grains $a$ with charge $Ze$, $J_i(Z)$ is the rate at which the ions arrive at a grain with size $a$ and charge $Ze$, and $\theta(y)=0$ if $y<0$ and $\theta(y)=1$ if $y\ge 1$.
$J_i(Z)$ is given by
\begin{equation}
J_i(Z) = n(X^i) \Bigg( \frac{8 k T}{\pi m_p} \Bigg)^{1/2} A_X^{-1/2} \pi a^2 \tilde{J}( \tau_i, \xi_i)
\end{equation}
where $\tilde{J}$ is the reduced rate coefficient \citep[see][]{Draine87}.
For the photoelectric ejection rate we use the analytical fit given in \citet{Bakes94}
\begin{equation}
J_\text{pe} = 2.5\times10^{-13} (13.6 - \text{IP}_z)^{1.983} N_\text{C} f_y(N_\text{C})
\end{equation}

\subsection{Reactions with vibrationally excited H$_2$}
As in \citet{Tielens85}, H$_2^*$ is treated as single species with an effective energy of 2.6 eV and corresponding to a vib-rotational pseudo-level $\nu =6$ of H$_2$. In this approximation H$_2^*$ is formed by FUV pumping of H$_2$ and destroyed by radiative and collisional decay by H and H$_2$. Following \citet{Tielens85}, the FUV pumping rate is 9 times the H$_2$ photodissociation rate and the radiative decay rate is $2\times 10^{-7}$ s$^{-1}$. For collisional de-excitation by collision with H and H$_2$ we use the rates given in Eq. A14 of \citet{Tielens85}. Reactions with vibrationally excited H$_2$ are taken into account for H, H$_2$, C$^+$, C, O, OH, CH and O$_2$ \citep{Tielens85}.

\section{Grain surface chemistry}
\label{sec:grain_chemistry}
Dust grain properties are one of the main inputs to the gas grain chemical model and are passed on from the thermochemical model results along with the disk structure; these include the integrated dust cross sectional area $\sum \pi a(i)^2n_d (i,l)$ where $i$ and $l$ represent the grain size ($a(i)$) bins  and compositions respectively. The zeroth and first order moments of the dust grain density distribution needed to calculate the growth of a grain due to condensation (see \S 2.3) are also inputs obtained from the thermochemical model. Grains of different sizes and compositions have different temperatures because of their optical properties and the dust temperature $T_d$ is a function of spatial location $(r,z)$, the grain size and its composition. The gas grain chemical code, howeover, uses a single temperature for computational efficiency. This is determined from the thermochemical modeling, and $T_d$ used is a cross-section and density weighted dust temperature. In effect, this is the temperature that the gas would attain if dust collisions were the only factor in thermal balance---warm grains collide to heat gas molecules while colder grains collide to cool it. We consider the following types of chemical reactions on grain surfaces:

\subsection{Accretion}

The adsorption rate coefficient of gas phase molecules onto the grain surface is given by

\begin{equation}
k_\text{acc}(i) = \alpha_i v_i \langle \sigma_d n_d \rangle
\end{equation}
where $\alpha_i$ is the sticking probability of species $i$, $v_i$ its thermal speed and $\langle \sigma_d n_d \rangle$ the product of the dust cross sectional area and the dust density averaged over the grain size distribution defined in Eq. \ref{eq:geo_cross_sec2}.
Sticking probabilities are assumed to be unity for all the neutral species, except H and H$_2$ for which temperature dependent expressions derived from laboratory experiments are used (see \citet{Matar10} and \citet{Chaabouni12}, for more information).

\subsection{Diffusion}

Diffusion at the surface and in the bulk of the ice is assumed to be driven by thermal diffusion. For both surface and mantle species, the time scale associated with this process is
\begin{equation}\label{eq:thop}
t_\text{hop}(i) = \nu_i^{-1} \exp\Bigg(\frac{E_\text{diff}(i)}{T_\text{d}}\Bigg)
\end{equation}
where $T_d$ is the dust temperature, $ \nu_i=\sqrt{2N_\text{sites} k E_\text{bind}(i)/ \pi^2 m_i}$ is the vibrational frequency of the species $i$, $E_\text{bind}$ its binding energy, $m_i$ its mass and $E_\text{diff}$ the barrier against diffusion, and the number of adsorption sites per cm$^2$ $N_\text{sites} = 1.0/d_\text{sites}^2 \sim 10^{15}$ cm$^{-2}$. Diffusion energy barriers are in general poorly constrained and only a handful of theoretical and experimental data are available. Therefore, $E_\text{diff}$ is usually defined as a fixed fraction of the binding energy $E_\text{bind}$ for each chemical species. Adopted values for $E_\text{diff}/E_\text{bind}$ vary greatly and range between 0.3 \citep{Hasegawa92} to 0.8 \citep{Ruffle00}. Experimental results of \citet{Katz99} and \citet{Perets05} on the formation of H$_2$ on grains and for which both $E_\text{diff}$ and $E_\text{bind}$ have been estimated lead to $E_\text{diff}/E_\text{bind}=0.8$ . More recently \citet{Hama12} estimated the energy diffusion barrier of atomic hydrogen to be $E_\text{diff}\sim250$K, which gives $E_\text{diff}/E_\text{bind}\sim0.4$ if $E_\text{bind}(\text{H})=650$K \citep{AlHalabi07} is assumed. Even though there is no fundamental reason for having a fixed ratio between different chemical species, this ratio is also supported by more recent theoretical work on CO and CO$_2$ for which $E_\text{diff}/E_\text{bind}\sim0.3$ and 0.4 on water ice \citep{Karssemeijer14} respectively. In our case we use two efficiencies depending upon the molecule is at the surface or in the mantle. For surface species we use $E^s_\text{diff}/E_\text{bind}=0.4$ and assume $E^m_\text{diff}/E_\text{bind}=0.8$ for mantle species (the diffusion energy barrier in the mantle is decreased by a factor of two as compared to the surface). In the mantle, we assume that diffusion is the result of a collective moleular motion rather than an individual process and thus set the diffusion energies of all the species having $E^m_\text{diff}<E^m_\text{diff}(\rm H_2O)$ to that of water (i.e. the main ice constituent). Exceptions are for H, H$_2$, C, N and O which are considered to be light enough to diffuse in the bulk of the ice with their own diffusion energy. Finally, energy barriers for swapping between surface layers and deeper ice manyle layers are assumed to be equal to the diffusion energy barriers in the mantle.

\subsection{Reaction}

Reactions at the surface are assumed to proceed via the Langmuir-Hinshelwood mechanism, in which pre-adsorbed molecules diffuse over the surface and possibly react when they meet. The reaction rate for this mechanism is given by
\begin{equation}
\label{eq:reac}
    R^s_{lj}(i) = \kappa_{lj}\Bigg[\frac{1}{t^s_{\text{hop},l}} + \frac{1}{t^s_{\text{hop},j}}\Bigg] \frac{n_\text{s}(l) n_\text{s}(j)}{n_\text{sites}}
\end{equation}
where $\kappa_{lj}$ is the probability that the reaction occurs, and $n_\text{sites}$  the  number density of adsorption sites defined  in Eq. \ref{eq:nsites}. For mantle reactions, this expression is reduced by a factor 
\begin{equation}
N_\text{lay}^m = \frac{\sum n_m(i)}{n_\text{sites}} 
\end{equation}
where $N_\text{lay}^m$ the total number of mantle layers on the grain. This modification accounts for the fact that the time needed for a mantle species to scan the entire grain mantle increases as the mantle grows.

$\kappa_{lj}$ is assumed to be unity for all barrier less reactions. For reactions with barrier we compute $\kappa_{lj}$ as the result of the competition among reaction and diffusion \citep[see][]{Ruaud16}
\begin{equation}
    \kappa_{lj} = \frac{\max(\nu_{l},\nu_{j})\kappa_{lj,0}}{\max(\nu_{l},\nu_{j})\kappa_{lj,0} + k_{\text{hop},l} + k_{\text{hop},j} }
\end{equation}
where $\kappa_{lj,0}$ is expressed as the $\exp(-\Delta E_{lj}/ T_\text{d})$ with $\Delta E_{lj}$ the activation barrier of the reaction. For reactions involving H and H$_2$ we use the quantum mechanical probability for tunneling through a rectangular barrier of thickness $a$; $\kappa_{lj,0}=\exp[-2(a/ \hbar)(2\mu \Delta E_{lj})^{1/2}]$, with $\mu$ the reduced mass of the system.

\subsection{Photodissociation}
  
Photodissociation rates for ices are generally not known and only a handful of photodissociation cross sections have been obtained experimentally \citep[see][for a review on photochemistry in ices]{Oberg16}. Photodissociation rates for ices in gas-grain chemical models are therefore generally assumed to be similar to the gas phase. We follow this approach and consider that the photodissociation rates of ice species are similar to their gas-phase counterpart (see Appendix \ref{sec:photoprocesses}). For mantle species, we take into that the absorption of the FUV photons in the upper layers of the ice results in the shielding of the lower layers. In this approach, the mantle photodissociation rates are computed using a photo absorption probability per monolayer $P_\text{abs}=0.007$ \citep{Andersson08}. The photodissociation rate coefficient of a species $i$ buried in the $j$-th layer is thus
\begin{eqnarray}
k^j_\text{ph,ice}(i) &=& k_\text{ph}(i) (1-P_\text{abs})^{j-1}
\end{eqnarray}
where $k_\text{ph}(i)$ is given by Eq. \ref{eq:gas_photo_sig} and Eq. \ref{eq:gas_photo}. The total photodissociation rate of species $i$ is obtained by multiplying by the number density of species $i$ in each monolayer $j$ and summing over the number of monolayers $N_\text{lay}$
\begin{eqnarray}
R_\text{ph,ice}(i) &=& \sum_{j=1}^{N_\text{lay}} k^j_\text{ph,ice}(i) n^j_\text{ice}(i)
\end{eqnarray}
This expression takes into account a global attenuation of the FUV photons due to the material present along the line of sight (see Appendix \ref{sec:photoprocesses}) as well as a local attenuation in the mantle by absorption of FUV photons by the upper layers of the ice.

\subsection{Desorption}

Desorption from the surface of the grain includes thermal desorption, photodesorption and chemical desorption.

\subsubsection{Thermal desorption}
The rate coefficients for thermal desorption are given by
\begin{equation}
k_\text{des,td}(i) = \nu_i \exp\Bigg(-\frac{E_\text{bind}(i)}{T_d}\Bigg)
\end{equation}
where $\nu_i$ is similar to Eq. \ref{eq:thop}.

\subsubsection{Photodesorption}
The rate coefficients for photodesorption are calculated following 
\begin{equation}
k_\text{des,pd}(i) = Y_\text{pd}(i) F_\text{FUV} \frac{\langle \sigma_d n_d \rangle}{ n_\text{sites}}
\end{equation}
where $F_\text{FUV}$ is the incident FUV photon flux on the grain and $Y_\text{pd}$ the photodesorption yield of the species $i$. Over the past few years, quantitative photodesorption yields have been measured experimentally for pure and mixed ices composed of molecules of particular interest using broadband and tunable FUV sources \citep[see][and references therin]{Ruaud16}. Experiments conducted using mixed ices show that photodesorption can be the result of a transfer of energy from the absorbing molecule to a neighboring one that gets desorbed. The indirect nature of this process makes photodesorption yields strongly dependent on the composition of the ice and they can vary by more than an order of magnitude \citep[see][on the CO-N$_2$ system for instance]{Bertin13}.  As in \citet{Ruaud16} we therefore use a simplistic approach and consider a single photodesorption yield of $Y_\text{pd}=1.0\times10^{-4}$ for all the molecules rather than individual ones determined from experiments. 

\subsubsection{Chemical desorption}
Chemical desorption is a process by which the product of an exothermic reaction is desorbed from the grains due to the excess energy. In our model we use the formalism proposed by \citet{Garrod07}, based on Rice-Ramsperger-Kessel theory which relates the excess energy and the binding energy of the newly formed molecule to a desorption probability
\begin{equation}
f_\text{RD} = \frac{a P}{1+aP}~\text{with} ~ P=\Bigg( 1-\frac{E_\text{bind}}{\Delta H_\text{R}}\Bigg)^{s-1}
\end{equation}
where $a$ is the efficiency of the process. We assume $a=0.01$, although this efficiency is not well constrained and can vary between $0.01$ and $0.1$ in the literature \citep{Garrod07,Vasyunin13,Ruaud16}. $P$ is the probability of desorption which depends on the exothermicity of the reaction ($\Delta H_\text{R}$), the binding energy of the product ($E_\text{bind}$) and the number of vibrational modes ($s$) in the molecule/surface-bond system \citep[see][]{Garrod07}. $s=2$ for diatomic species; for all the others $s=3N-5$ with $N$ the number of atoms in the molecule. 

\subsection{Formation of OCS by $\rm CO + HS$ on grains}
\label{ocs_form}
The reaction between sCO and sHS is usually not taken into account in chemical models considering its high endothermicity in the gas-phase $\Delta E = 6200$ K.
However this reaction has been proposed as a possible reaction pathway to sOCS by \citet{Jimenez14} and \citet{Chen15} who have conducted laboratory irradiation experiments on ices containing a mixture of sH$_2$S and sCO as well as sH$_2$S and sCO$_2$. This reaction pathway has been investigated theoretically on coronene by \citet{Adriaens10}. In this study, the authors show that the most favorable outcome for this reaction is the formation of $trans$-HSCO which only requires an activation barrier of 3.3kJ/mol (i.e.\  $\Delta E \sim 400$ K). The authors propose that $trans$-HSCO can be further hydrogenated to form $\rm sOCS + sH_2$ or $\rm sH_2S + sCO$ depending on the angle of approach of the incoming hydrogen. In our case we consider that the reaction $\rm sCO + sHS$ proceeds with an activation barrier of 400 K and consider $90\%$ sHSCO relaxation and $10\%$ formation of sOCS. For hydrogenation of sHSCO we consider that sH$_2$S and sOCS form with equal branching ratios. However we note that more theoretical work is needed to constrain this pathway on ice and to derive the different branching ratios of the system.

\bibliographystyle{aasjournal}
\bibliography{bibliography}

\end{document}